\documentclass[epj]{svjour}
\usepackage{cite,mcite}
\usepackage{graphicx}
\usepackage{epsfig}
\usepackage{subfigure}
\usepackage{atlasphysics}
\usepackage{amsmath}
\usepackage{amssymb}
\usepackage{hyperref}
\usepackage{mathptmx}
\usepackage{helvet}
\newcommand{\ignorenow}[1] { }

\newcommand{\pp}{\ensuremath{\, \it{pp}}}
\newcommand{\ppbar}{\ensuremath{\, \it{p\overline{p}}}}
\newcommand{\pythia}{{\textsf {\sc PYTHIA}}}
\newcommand{\phojet}{{\textsf {\sc PHOJET}}}
\newcommand{\sigmavis}{\ensuremath{\, \sigma_{vis}}}
\newcommand{\muvis}{\ensuremath{\mu^{vis}}}
\newcommand{\sigmainel}{\ensuremath{\, \sigma_{inel}}}
\newcommand{\mbtsone}{\textsf{\small MBTS\_Event\_OR}}
\newcommand{\mbtsoneone}{\textsf{\small MBTS\_Event\_AND}}
\newcommand{\mbtsOFF}{\textsf{\small MBTS\_Timing}}
\newcommand{\LARtiming}{\textsf{\small LAr\_Timing}}
\newcommand{\mbtspvtx}{\textsf{\small PrimVtx}}
\newcommand{\chpart}{\textsf{\small ChPart}}
\newcommand{\lucidEventOr}{\textsf{\small LUCID\_Event\_OR}}
\newcommand{\lucidEventAnd}{\textsf{\small LUCID\_Event\_AND}}
\newcommand{\lucidHitOr}{\textsf{\small LUCID\_Hit\_OR}}
\newcommand{\lucidHitAnd}{\textsf{\small LUCID\_Hit\_AND}}
\newcommand{\lucidZeroOr}{\textsf{\small LUCID\_Zero\_OR}}
\newcommand{\lucidZeroAnd}{\textsf{\small LUCID\_Zero\_AND}}
\newcommand{\bcmEventOr}{\textsf{\small BCM\_Event\_OR}}
\newcommand{\bcmEventAnd}{\textsf{\small BCM\_Event\_AND}}
\newcommand{\bcmEventXOr}{\textsf{\small BCM\_Event\_XORC}}

\newcommand{\zdcEventAnd}{\textsf{\small ZDC\_Event\_AND}}
\newcommand{\zdcA}{\textsf{\small ZDC\_A}}
\newcommand{\zdcC}{\textsf{\small ZDC\_C}}
\newcommand{\eventAnd}{\textsf{\small Event\_AND}}
\newcommand{\tinyeventAnd}{\textsf{\tiny Event\_AND}}
\newcommand{\eventOr}{\textsf{\small Event\_OR}}
\newcommand{\tinyeventOr}{\textsf{\tiny Event\_OR}}

\newcommand{\zeroOr}{\textsf{\small Zero\_OR}}
\newcommand{\smmbtsone}{\textsf{\small MBTS\_OR}}
\newcommand{\smmbtsoneone}{\textsf{\small MBTS\_AND}}
\newcommand{\smmbtsOFF}{\textsf{\small MBTS\_timing\_Event}}
\newcommand{\smLARtiming}{\textsf{\small LAr\_timing\_Event}}
\newcommand{\smmbtspvtx}{\textsf{\small PrimVtx\_Event}}
\newcommand{\smchpart}{\textsf{\small ChPart\_Event}}
\newcommand{\smlucidEventOr}{\textsf{\small LUCID\_Event\_OR}}
\newcommand{\tinylucidEventOr}{\textsf{\tiny LUCID\_Event\_OR}}
\newcommand{\smlucidEventAnd}{\textsf{\small LUCID\_Event\_AND}}

\newcommand{\smbcmEventOr}{\textsf{\small BCM\_Event\_OR}}

\newcommand{\smzdcA}{\textsf{\small ZDC\_A}}
\newcommand{\smzdcC}{\textsf{\small ZDC\_C}}
\newcommand{\vdM}{{\it vdM}}

\def\thedate{\today}
\begin{document}
\hugehead
\title{Luminosity Determination in \pp\ Collisions at $\sqrt s = 7$ TeV 
Using the ATLAS Detector at the LHC}
\author{The ATLAS Collaboration}
\date{\thedate}
\institute{}
% The abstract
%
\abstract{
Measurements of luminosity obtained using the ATLAS detector
during early running of the 
Large Hadron Collider (LHC) at $\sqrt s = 7$ TeV 
are presented.
The luminosity is independently determined using several detectors and
multiple algorithms, each having different acceptances, 
systematic uncertainties and sensitivity to background.
The ratios of the luminosities obtained from these methods
are monitored as a function of time and of $\mu$, the average number
of inelastic interactions per bunch crossing. Residual time- and $\mu$-dependence
between the methods is less than 2\%
for $0<\mu<2.5$.
Absolute luminosity calibrations, performed using beam separation scans,
have a common systematic uncertainty of $\pm 11\%$, dominated by 
the measurement of the LHC beam currents.  
After calibration, the luminosities obtained from the different methods
differ by at most $\pm 2\%$.
The visible cross sections
measured using the beam scans are compared to predictions obtained with
the \pythia\ and \phojet\ event generators and the ATLAS detector
simulation.
}
 % end of PACS codes               

\maketitle
%
%%%%%%%%%%%%%%%%%%%%%%%%%%%%%%%%%%%%%%%%%%%%%%%%%%%%%%%%%%%%%%%%%%%%%%%%%%%%%%%
% Introduction
%%%%%%%%%%%%%%%%%%%%%%%%%%%%%%%%%%%%%%%%%%%%%%%%%%%%%%%%%%%%%%%%%%%%%%%%%%%%%%%
%
\section{Introduction and Overview}
A major goal of the ATLAS\cite{bib:ATLASDetectorPaper} physics program for 
2010 is the measurement of cross sections for Standard Model processes.  
Accurate determination of  the luminosity is an essential ingredient of this 
program. This article describes the first results on luminosity determination, 
including an assessment of the systematic uncertainties, for data taken at the 
LHC\cite{bib:LHC} in proton-proton collisions at a centre-of-mass energy 
$\sqrt{s}=7$~TeV. It is organized as follows.

\par
The ATLAS strategy for measuring and calibrating the luminosity  is outlined  
below and is followed in Section~\ref{sec:detectorDescription} 
by a brief description of the subdetectors 
used for luminosity determination. Each 
of these detectors is associated with one or more luminosity algorithms,
described in Section~\ref{sec:methods}. 
The absolute calibration of these algorithms using
beam-separation scans forms the subject of Section~\ref{sec:Vdm}. The internal 
consistency of the luminosity measurements is assessed in 
Section~\ref{sec:Cnstcy}. Finally, the scan-based calibrations are compared
in Section~\ref{sec:MC} to those predicted using the \pythia\cite{bib:pythia} 
and \phojet\cite{bib:phojet} event 
generators coupled to a full GEANT4\cite{bib:G4} simulation of the ATLAS detector response\cite{bib:ATLASSimPaper}.
Conclusions are summarized in Section~\ref{sec:conclusions}.

\label{sec:overview}
The  luminosity of a \pp\ collider can be expressed as
\begin{equation}
\mathcal{L} = \frac{R_{inel}}{\sigma_{inel}}
\end{equation}
where  $R_{inel}$ is the rate of inelastic collisions and $\sigma_{inel}$ is
the \pp\ inelastic cross section.  If a collider operates at a revolution
frequency $f_r$ and $n_b$ bunches cross at the interaction point, 
this expression can be rewritten as
\begin{equation}
\mathcal{L} = \frac{{\mu n_b f_r }}{{\sigma _{inel} }}
\label{eq:lumiToMu}
\end{equation}
where $\mu $ is the average number of inelastic interactions per 
bunch crossing (BC). Thus, the instantaneous luminosity can be determined
using any method that measures the ratio $\mu/\sigma_{inel}$.

\par
A fundamental ingredient of the ATLAS strategy to assess and control the 
systematic uncertainties affecting the absolute luminosity determination is to 
compare the measurements of several luminosity detectors, most of which use 
more than one counting technique. These multiple detectors and 
algorithms are characterized by 
significantly different acceptance, response to pile-up (multiple \pp\
interactions within the same bunch crossing), 
and sensitivity to 
instrumental effects and to beam-induced backgrounds. The level of consistency 
across the various methods, over the full range of single-bunch luminosities and 
beam conditions, provides valuable cross-checks as well as an estimate of the 
detector-related systematic uncertainties.

\par
Techniques for luminosity determination can be classified as follows:
\begin{itemize}
\item {\it Event Counting:}  here one determines the fraction of bunch crossings 
during which a specified detector registers an \\
``event'' satisfying a 
given selection requirement. 
For instance, a bunch crossing can be said to contain an ``event'' if 
at least one \pp\ interaction in that crossing induces at
least one observed hit in the detector being considered.
\item {\it Hit Counting:}  here one counts the number of hits (for
example,  electronic  channels or energy clusters above a specified threshold) 
per bunch crossing in a given detector.
\item {\it Particle Counting:} here one determines the distribution of the 
number of particles per beam crossing (or its mean) inferred 
from reconstructed quantities ({\it e.g.} 
tracks), from pulse-height distributions or from other observables 
that reflect the instantaneous particle flux traversing the detector ({\it e.g.} 
the total ionization current drawn by a liquid-argon calorimeter sector).  
\end{itemize}

\par
At present, ATLAS relies 
only
on event-counting methods for the determination of the absolute luminosity.
Equation~\ref{eq:lumiToMu} can be rewritten as:
\begin{equation}
\mathcal{L} = 
\frac{{\mu n_b f_r }}{{\sigma _{inel} }} =
\frac{{\mu ^{vis} n_b f_r }}{{\varepsilon \sigma _{inel} }} =
\frac{{\mu ^{vis} n_b f_r }}{{\sigma _{vis} }}
\label{eq:calib}
\end{equation}
where $\varepsilon$ is the efficiency for one inelastic \pp\ collision to 
satisfy the event-selection criteria, and $\mu^{vis} \equiv \varepsilon \mu$ is 
the average number of visible inelastic interactions per BC ({\it i.e.} the mean 
number of \pp\ collisions per BC that pass that ``event'' selection). The 
visible cross section $\sigma_{vis}\equiv \varepsilon \sigma_{inel}$ is the 
calibration constant that relates the measurable quantity $\mu^{vis}$ to the 
luminosity $\mathcal{L}$. Both $\varepsilon$ and $\sigma_{vis}$ depend on the 
pseudorapidity distribution and particle composition of the collision products, 
and are therefore different for each luminosity detector and algorithm.

\par
In the limit $\mu_{vis} \muchless 1$, the average number of visible inelastic 
interactions per BC is given by the intuitive expression
\begin{equation}
\mu^{vis} \approx \frac{N}{N_{BC}}
\label{eq:lowmuApprox} 
\end{equation}
where $N$ is the number of events passing the selection criteria that are 
observed during a given time interval, and $N_{BC}$ is the number of bunch 
crossings in that same interval. When $\mu$ increases, the probability that two 
or more \pp\ interactions occur in the same bunch crossing is no longer 
negligible, and \muvis\ is no longer linearly related to the raw event count 
$N$. Instead \muvis\ must be calculated taking into account Poisson statistics, 
and in some cases, instrumental or pile-up related effects 
(Section~\ref{subsec:mudependence}). 

\par
Several methods can be used to determine \sigmavis .
At the Tevatron, luminosity measurements are normalized to the
total inelastic \ppbar\ cross section, with simulated
data used to determine the event- or hit-counting
efficiencies\cite{bib:vaia,bib:Fermilab-FN-074}.  
Unlike the case of the Tevatron, where the \ppbar\ cross
section was determined 
\footnote{ In fact, Tevatron cross sections were measured at $\sqrt s=1.8$
TeV and extrapolated to  $\sqrt s = 1.96$ TeV.}
independently by two experiments,
the \pp\ inelastic cross section at 7 TeV has not been measured yet. 
Extrapolations from lower energy involve significant systematic uncertainties,
as does the determination of $\varepsilon$, which depends on the
modeling of particle momentum distributions and multiplicity 
for the full \pp\ inelastic cross section.
In the future, the ALFA detector\cite{bib:ALFA}
will provide an absolute luminosity calibration at ATLAS through
the measurement of elastic \pp\-scattering at small angles in the 
Coulomb-Nuclear Interference  region.  
In addition, it is possible to normalize cross section measurements
to  electroweak processes for which
precise NNLO calculations exist, for example $W$ and $Z$ production\cite{bib:NNLO}.
Although the cross section for the production of electroweak bosons in \pp\
collisions at $\sqrt{s}=7$~TeV has been measured
by ATLAS \cite{bib:ATLASWZ} and found to be in agreement with the
Standard Model expectation, with 
experimental and theoretical
systematic uncertainties of $\sim 7\%$, we choose not to use these
data as a luminosity calibration, since such use would preclude future
comparisons with theory.  
%However,
%the increased instantaneous luminosity expected in the future should
However, in the future, it will be possible to monitor the variation of 
luminosity with time
using $W$ and $Z$ production rates.

\par
An alternative is to calibrate the 
%%%  event- and hit-
counting techniques using the 
absolute luminosity $\mathcal L$ inferred from measured accelerator 
parameters~\cite{bib:helmut}~\cite{bib:Kozanecki:2008zz}:
\begin{equation}
{\mathcal L} = \frac{{n_b f_r n_1 n_2 }}{{2\pi \Sigma _x \Sigma _y
}}\label{defLumi}
\end{equation}
where $n_1$ and $n_2$ are the numbers of particles in the two colliding bunches 
and $\Sigma _x$ and $\Sigma _y$ characterize 
the widths of the horizontal and vertical beam 
profiles. One typically measures $\Sigma _x$ and $\Sigma_y$ using {\it van der 
Meer (vdM)} scans (sometimes also called {\it beam-separation} or {\it 
luminosity} scans)\cite{bib:vdm}. The observed event rate
%%%  or hit-rate 
is recorded 
while scanning the two beams across each other first in the horizontal ($x$), 
then in the vertical ($y$) direction. This measurement yields two bell-shaped 
curves, with the maximum rate at zero separation, from which one extracts 
the  values of $\Sigma _x$ and $\Sigma _y$ (Section~\ref{sec:Vdm}). 
The luminosity at zero separation can then 
be computed using Equation~\ref{defLumi}, and \sigmavis\ extracted from 
Equation~\ref{eq:calib} using the measured values of $\mathcal L$ and \muvis.

\par
The {\it vdM} technique allows the determination of \sigmavis\ without 
{\it a priori} knowledge of the inelastic \pp\ cross section or of detector 
efficiencies.  Scan results can therefore be used to test
the reliability of Monte Carlo event generators and of the ATLAS simulation
by comparing the visible cross sections predicted by the Monte Carlo 
for various detectors and algorithms to those obtained from the scan data.

\par
ATLAS uses the {\it vdM} method to obtain its absolute luminosity calibration
both for online monitoring and for offline analysis. Online, the luminosity at 
the ATLAS interaction point (IP1) is determined approximately once per second using 
the counting rates from the 
detectors and algorithms described in Sections~\ref{sec:detectorDescription} and 
\ref{sec:methods}. The raw event 
%%%  (or hit) 
count $N$ is converted to a visible 
average number of interactions per crossing $\mu_{vis}$ as described in 
Section~\ref{subsec:mudependence}, and expressed as an absolute luminosity using 
the visible cross sections $\sigma_{vis}$  measured during beam-separation scans. 
The results of all the methods are displayed in the ATLAS control room, 
and the luminosity from a single online ``preferred'' algorithm is transmitted 
to the LHC control room, providing real-time feedback for accelerator tuning.  

The basic time unit for storing luminosity information for later use
is the {\em Luminosity Block} (LB).  The duration of a LB is approximately
two minutes, with begin and end times set by the ATLAS data acquisition
system (DAQ).  All data-quality information, as well as the luminosity, are
stored in a relational database for each LB.  
The luminosity tables in the offline database allow for storage of multiple
methods for luminosity determination
and are versioned so that updated calibration constants
can be applied.  The results of all online
luminosity methods are stored, and
results from additional offline algorithms are added.  
This infrastructure enables comparison of the results from different
methods as a function of time.  After data quality checks have been
performed and calibrations have been validated, one algorithm is chosen as the
``preferred'' offline algorithm for physics analysis and
stored as such in the database.  
Luminosity information is stored as delivered luminosity. Corrections for 
trigger prescales, DAQ deadtime and other sources of data loss are
performed on an LB-by-LB basis when the integrated luminosity is calculated.

\section{The ATLAS Luminosity Detectors}
\label{sec:detectorDescription}
The ATLAS detector is described in detail in
Ref.~\cite{bib:ATLASDetectorPaper}.  This section provides a brief
description of the subsystems used for luminosity measurements, 
arranged in order of increasing pseudorapidity.\footnote{
ATLAS uses a coordinate system where the nominal interaction point is
at the centre of the detector. The direction of beam 2 (counterclockwise 
around the LHC ring) defines the $z$-axis; the $x$-$y$ plane is transverse
to the beam. The positive $x$-axis is 
defined as pointing to the centre of the ring, and the positive $y$-axis
upwards.  Side-A of the detector is on the-positive $z$ side and
side-C on the negative-$z$ side. The azimuthal angle $\phi$ is 
measured around the beam axis.  The pseudorapidity $\eta$ is defined
as $\eta = -\ln(\tan \theta/2)$ where $\theta$ is the polar angle 
from the beam axis.}
A summary of the relevant characteristics of these detectors is given
in Table~\ref{tab:detectors}.
\begin{table}
\begin{center}
\begin{tabular}{|l|c|c|l|}
\hline
Detector & Pseudorapidity Coverage & \#\ Readout Channels \\
\hline
Pixel & $|\eta|<2.5$ & $8\times 10^7$  \\
SCT & $|\eta|<2.5$ &$6.3\times 10^6$  \\
TRT & $|\eta|<2.0$ & $3\times 10^5$ \\
MBTS  &  $2.09<|\eta|< 3.84$  & 32 \\
LAr: EMEC  & $2.5<|\eta|< 3.2$ & $3\times 10^4$ \\
LAr: FCal &  $3.1<|\eta|< 4.9$   & 5632 \\
BCM &  $|\eta|=4.2$ & 8 \\
LUCID & $5.6<|\eta|< 6.0$ & 32 \\
ZDC &  $|\eta|>8.3 $ & 16  \\
\hline
\end{tabular}
\end{center}
\caption{Summary of relevant characteristics of the detectors used
for luminosity measurements.  For the ZDC, the number of readout 
channels only includes those used by the luminosity algorithms.}
\label{tab:detectors}
\end{table}

The Inner Detector is used to measure the momentum of charged particles.
It consists of three subsystems: a pixel detector, a silicon strip tracker
(SCT) and a transition radiation straw tube tracker (TRT).  These
detectors are located
inside a solenoidal magnet that provides a 2~T axial field.  The tracking
efficiency as a function of transverse momentum ($p_T$), averaged over
all pseudorapidity, rises from \hbox{$\sim 10\%$ } at 100 MeV to $\sim 86$\%
for $p_T$ above a few GeV\cite{bib:CONF046}.

For the initial running period at low instantaneous luminosity ($< 10^{33}~{\rm
cm}^{-2}{\rm s}^{-1}$), ATLAS has been equipped with segmented
scintillator counters, the 
Minimum Bias Trigger Scintillators (MBTS), located at $z=\pm365$~cm from the
collision centre.
The main purpose of the MBTS is to provide a trigger on 
minimum collision activity during a \pp\  bunch crossing.
Light emitted by the scintillators is
collected by wavelength-shifting optical fibers and guided to a
photomultiplier tube\\
(PMT).  The MBTS signals, after being shaped
and amplified, are fed into leading-edge
discriminators and sent to the central trigger processor (CTP).  
An MBTS hit is defined as a signal above the discriminator threshold ($50$~mV).

The precise timing ($\sim 1$~ns) provided by the liquid argon (LAr) calor\-imeter
is used to count events with collisions, therefore providing a measurement of
the luminosity. The LAr calor-imeter covers the region $|\eta|<4.9$.
It consists of 
the electromagnetic (EM) for $|\eta|<3.2$, the Hadronic Endcap for
$1.5 < |\eta|<3.2$
and the
Forward Calorimeter (FCal) for $3.1<|\eta|<4.9$. 
The luminosity analysis is
based on energy deposits in the Inner Wheel of the electromagnetic endcap
(EMEC) and the first layer of the FCal. The precise timing is used to
reject  background for the offline measurement of the luminosity.

The primary purpose of the Beam Conditions Monitor \\
(BCM)\cite{bib:BCM} is to monitor
beam losses and provide fast feedback to the accelerator operations team.
It is an essential ingredient of the detector protection system, 
providing a fast accelerator abort signal in the event of large beam loss.
The BCM consists of two arms of diamond sensors located at $z=\pm 184$~cm
and $r=5.5$~cm and uses programable front-end electronics (FPGAs) 
to histogram the single-sided and coincidence 
rates as a function of Bunch Crossing Identifier (BCID).  
These histograms are read out by the BCM
monitoring software and made available to other online applications through
the online network.  Thus, bunch-by-bunch rates are available and 
are not subject to DAQ deadtime.  The detector's value as
a luminosity monitor is further enhanced by its excellent 
timing ($~0.7$~ns) which
allows for rejection of backgrounds from beam-halo.

LUCID is a Cherenkov detector specifically designed for measuring the 
luminosity in ATLAS. 
Sixteen optically reflecting aluminum tubes filled 
with ${\rm C}_4{\rm F}_{10}$ gas surround 
the beampipe on each side of the interaction point.
Cerenkov photons created by charged particles in the gas are reflected by the 
tube walls until they reach PMTs situated at the back end of the 
tubes. 
The Cherenkov light created in the gas typically produces 60-70 photoelectrons,
while the quartz window adds another 40 photoelectrons to the signal. After 
amplification, the signals are split three-fold and presented to a set of 
constant fraction discriminators (CFDs), charge-to-digital converters  and 
32-bit flash ADCs  with 80 samplings. If the signal has a pulse height 
larger than the discriminator threshold (which is equivalent to 15 
photoelectrons) a tube is ``hit.'' The hit-pattern produced by 
all the discriminators is sent to a custom-built electronics card (LUMAT) 
which contains FPGAs that can be programmed with different luminosity 
algorithms. 
LUMAT receives timing signals from the LHC clock used
for synchronizing all detectors and
counts the  number of events or hits passing each luminosity algorithm
for each BCID  in an orbit.  It also 
records the number of orbits made by the protons in the LHC during
the counting interval. At present there are four algorithms implemented 
in the LUMAT firmware (see Section~\ref{sec:LUCID-methods}).
The data from  LUMAT  
are broadcast to the ATLAS online network
and archived for later offline use.
In addition, LUMAT
provides triggers for the CTP and sends the hit-patterns
to the DAQ. 
The LUCID 
electronics is decoupled from the DAQ  so that it 
can provide an online luminosity determination even if no global 
ATLAS run is in progress.

The primary purpose of the Zero-Degree Calorimeter (ZDC) is to detect 
forward neutrons and photons with $|\eta|> 8.3$ in both \pp\ and 
heavy-ion collisions.  
The ZDC consists of two arms located at $z=\pm 140 $~m  
in slots in the LHC TAN
(Target Absorber Neutral)\cite{bib:LHC}, occupying space that would otherwise
contain inert copper shielding bars.  In its final configuration, each
arm consists of  calorimeter
modules, one electromagnetic (EM) module (about 29
radiation lengths deep) followed by three hadronic modules (each about 1.14 
interaction lengths deep).   
The modules are composed of tungsten with an embedded matrix of quartz rods 
which are coupled to photo multiplier tubes and read out through CFDs.
Until July 2010
only the three hadronic modules were installed to allow running of the
LHCf experiment\cite{bib:lhcf}, which occupied the location where the EM
module currently sits.  
Taking into account the limiting
aperture of the beamline, the effective ZDC acceptance for neutrals
corresponds to 1~GeV in $p_T$ for a 3.5~TeV neutron or photon. 
Charged particles are swept out of the ZDC acceptance by 
the final-triplet quadrupoles;
Monte Carlo studies
have shown that neutral secondaries contribute
a negligible amount to the typical ZDC energy.  A hit in the ZDC is
defined as an energy deposit above CFD threshold.  The ZDC is 
fully efficient for energies above $\sim 400$ GeV. 

\section{Luminosity Algorithms}
\label{sec:methods}
The time structure of the LHC beams and its consequences for the luminosity 
measurement (Section~\ref{subsec:bunchbybunch}) drive the architecture of the 
online luminosity infrastructure and algorithms (Section~\ref{subsec:onlineAlgs}). 
Some approaches to luminosity determination, however, are only possible offline 
(Section~\ref{subsec:offlineAlgs}). In all cases, dealing properly with pile-up 
dependent effects (Section~\ref{subsec:mudependence}) is essential to ensure the 
precision of the luminosity measurements.

\subsection{Bunch Patterns and Luminosity Backgrounds}
\label{subsec:bunchbybunch}
The LHC beam is subdivided into 35640 RF-buckets of which nominally
every tenth can contain a bunch. Subtracting abort and injection gaps,
up to 2808 of these 3564 ``slots'', which are 25~ns long, can be filled
with beam.  Each of these possible crossings is labeled by an
integer BCID which is
%bunch crossing identifier (BCID) which is
stored as part of the ATLAS event record.

\par
\begin{figure*}
\subfigure[]{
\epsfig{
file=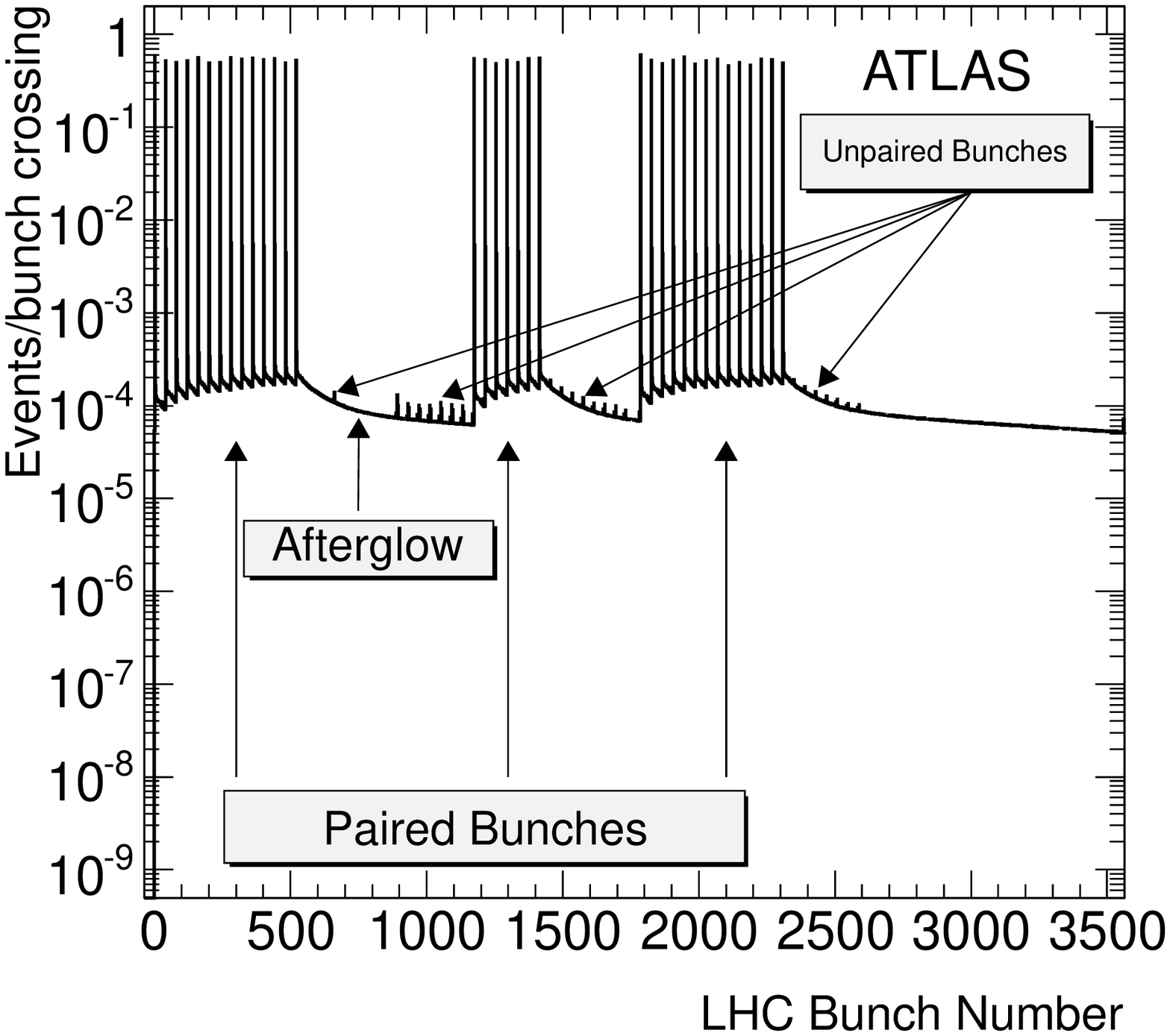,width=0.45\textwidth}
}
\subfigure[]{
\epsfig{
file=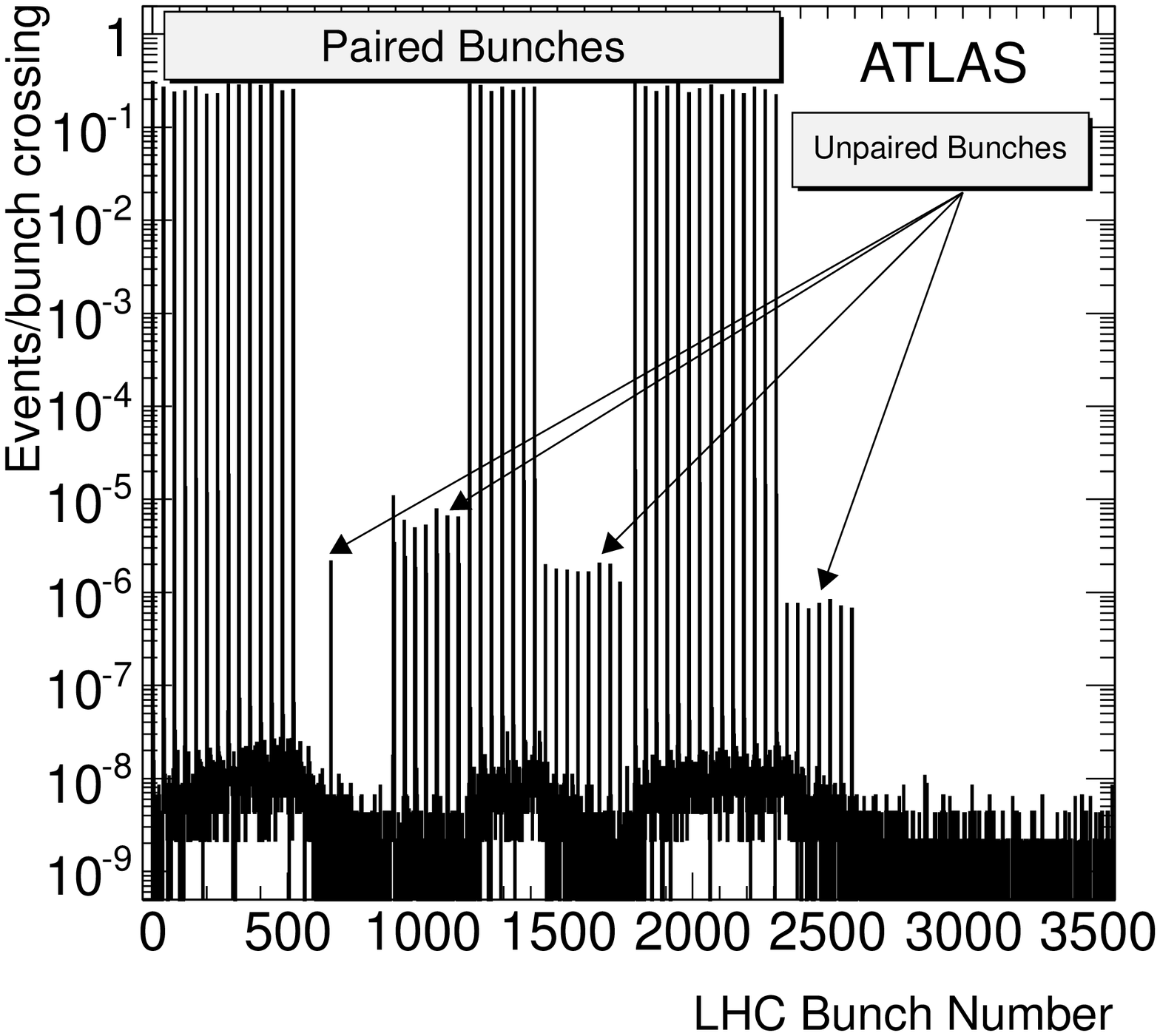,width=0.45\textwidth}
}
\caption{Bunch-by-bunch  event rate per bunch crossing in ATLAS run 
162882, as recorded by a LUCID algorithm that requires (a) at least one hit on 
either LUCID side (\eventOr), or (b) at least one hit on both LUCID sides
(\eventAnd) within the same BCID. 
}
\label{fig:LUCIDLumiPerBCID}
\end{figure*}

Figure~\ref{fig:LUCIDLumiPerBCID} 
displays the event rate per BC, as measured by two LUCID algorithms,
as a function of BCID and time-averaged over a run that lasted about 15 hours.
For this run, 35 bunch pairs collided in both ATLAS and CMS.  
These are called ``colliding'' (or 
``paired'') BCIDs.   Bunches that do not collide at IP1 are 
labelled ``unpaired.''
Unpaired bunches that undergo no collisions in any of the IPs are called
``isolated.''
The structures observed in this figure are visible 
in the bunch-by-bunch luminosity distributions of all the detectors discussed in 
this paper, although with magnitudes affected 
by different instrumental characteristics and background sensitivities. 
Comparisons of the event rates in colliding, unpaired, isolated and 
empty bunch crossings for different event-selection criteria provide 
information about 
the origin of the luminosity backgrounds, as well as quantitative estimates of the 
signal purity for each of these detectors and algorithms.

\par
Requiring at least one hit on at least one side (this is referred to as an \eventOr\ 
algorithm below) reveals a 
complex time structure (Fig.~\ref{fig:LUCIDLumiPerBCID}a). The colliding 
bunches are clearly distinguished, with a rate of about four orders of magnitude 
above background. They are followed by a long tail where the rate builds up when 
the paired BCID's follow each other in close succession, but decays slowly when no 
collisions occur for a sufficiently long time. This ``afterglow'' (which is also apparent 
when analyzing the luminosity response of \eventOr\ algorithms using the BCM or
MBTS) is dominated by slowly-decaying, low-energy radiation produced by \pp\ 
collision products that hit forward ATLAS components and scatter around the 
experimental cavern for tens of microseconds.
BCID's from unpaired and isolated bunches appear as small spikes above the 
afterglow background.  These spikes are the result of beam-gas and beam-halo 
interactions; in some cases, they may also contain a very small fraction of \pp\ 
collisions between an unpaired bunch in one beam and a satellite- or debunched-
proton component in the opposing beam.\footnote{In proton storage rings, a small 
fraction of the injected (or stored) beam may fail to be captured into (or may slowly 
diffuse out of) the intended RF bucket, generating a barely detectable unbunched 
beam component and/or coalescing into very low-intensity ``satellite'' bunches
that are separated from a nominal bunch by up to a few tens of buckets.} 

\par
For the \eventAnd\ algorithm (Fig.~\ref{fig:LUCIDLumiPerBCID}b), the 
coincidence requirement between the A- and C-sides suppresses the afterglow signal by 
an additional four orders of magnitude, clearly showing that this luminosity 
background is caused by random signals uncorrelated between the two sides. 
Unpaired-bunch rates for \lucidEventAnd\ lie 4-5 orders of magnitude lower than 
\pp\ collisions between paired bunches.

\par
This  figure illustrates several important points.  First, because only a fraction 
of the BCID's are filled, an algorithm that selects on colliding BCID's is 
significantly cleaner than one that is BCID-blind. Second, and provided only 
colliding BCID's 
are used, the background is small (LUCID) to moderate (MBTS) for \eventOr\ 
algorithms, and negligible for\\
 \eventAnd. In the \eventOr\ case, the background 
contains contributions both from afterglow and from beam-gas and beam-halo 
interactions: its level thus depends crucially on the time separation between 
colliding bunches.

\subsection{Online Algorithms}
\label{subsec:onlineAlgs}
\subsubsection{Online Luminosity Infrastructure}

Online luminosity monitoring and archiving can be made available even when only the 
core ATLAS DAQ infrastructure is active; this makes it possible 
to provide luminosity information for machine tuning independently of the ``busy'' 
state of the DAQ system and of the hardware status of most subdetectors (except for 
the CTP and for one or more of the luminosity detectors). In addition, since the 
online luminosity data are collected in the front-end electronics of each detector 
(or at the CTP input), there is no need for prescaling, even at the highest 
luminosities.

\par
The calculation and publication of instantaneous luminosities is performed by an 
application suite called the Online Luminosity Calculator (OLC). The task of the 
OLC is to retrieve the raw luminosity information (event or hit counts, number of 
colliding bunches $n_b$, and number of LHC orbits in the time interval considered)
from the online network and to use these data to determine $\mu$ and hence the 
measured luminosity. For each luminosity algorithm, the OLC outputs the 
instantaneous luminosity, averaged over all colliding BCIDs,
at about 1 Hz. These values are displayed on online monitors,
stored in the ATLAS online-monitoring archive and shipped to the 
LHC control room to assist in collision optimization at IP1. In addition, the OLC 
calculates the luminosity averaged over the current luminosity block (in 
all cases the 
luminosity averaged over all colliding BCIDs, 
and when available the bunch-by-bunch 
luminosity vector) and stores these in the ATLAS conditions database.

\par
Most methods provide an LB-averaged luminosity measured from colliding 
bunches only, but for different detectors the requirement is imposed at different 
stages of the analysis. The BCM readout driver and the LUCID LUMAT module 
provide bunch-by-bunch raw luminosity information for each LB, as well as the 
luminosity per LB summed over all colliding BCID's. For these two detectors, the 
OLC calculates the total ({\it i.e.} bunch-integrated) luminosity using an 
extension of Equation~\ref{eq:calib} that remains valid even when 
each bunch pair produces a different 
luminosity (reflecting a different value of $\mu$) because of 
different bunch currents and/or emittances:
\begin{equation}
{ \cal L} = \sum_{i \in BCID} \mu^{vis}_i \frac{f_r} {\sigma_{vis}}
\label{eq:sumBCIDlumi}
\end{equation}
where the sum is performed over the colliding BCID's. This makes it possible to 
properly apply the pile-up correction  
bunch-by-bunch (Section~\ref{subsec:mudependence}).

\par
For detectors where bunch-by-bunch luminosity is unavailable online, 
Equation~\ref{eq:calib} is used, with $\muvis$ computed using the known number of 
paired BCID's and the raw luminosity information averaged over either the colliding 
BCID's (this is the case for the MBTS) or all BCID's (the front-end luminosity 
infrastructure of the ZDC provides no bunch-by-bunch capability at this time).

\par
For the MBTS, which lacks appropriate FPGA capabilities in the front end, the 
selection of colliding bunches is done through the trigger system. The BCID's that 
correspond to colliding bunches are identified and grouped in a list called the 
``physics bunch group,'' which is used to gate the physics triggers.  A second set 
of triggers using unpaired bunches is used  offline to estimate beam backgrounds.  
The MBTS counters 
provide trigger signals to the CTP, which then uses bunch-group information to 
create separate triggers for physics and for unpaired bunch groups. The CTP scalers 
count the number of events that fire each trigger, as well as the number of LHC 
orbits (needed to compute the rate per bunch crossing). Every $10\,{\rm s}$ these 
scalers are read out and published to the online network. Three values are stored 
for each trigger type: trigger before prescale (TBP), trigger after prescale 
and trigger after veto (TAV).  The TBP counts are calculated directly using inputs 
to the CTP and are therefore free from any dead time or veto 
(except when the DAQ is paused),
 while the TAV corresponds to the rate of accepted events 
for which a trigger fired. To maximize the statistical power of the measurement and 
remain unaffected by prescale changes, online luminosity measurements by the MBTS 
algorithms use the TBP rates.

\subsubsection{BCM Algorithms}

Out of the four sensors
on each BCM side, only two are currently used for online luminosity
determination.
Three online algorithms, implemented in the firmware of the 
BCM readout driver, 
report results:
\begin{itemize}
\item \bcmEventOr\ counts the number of events per BC in which at least one hit above threshold occurs on either the A-side, the C-side or both, within a 
12.5~ns window centred on the arrival time of particles originating at IP1;
\item \bcmEventAnd\ counts the number of events per BC where at least one hit above 
threshold is observed, within a 12.5~ns-wide coincidence window, both on the A- 
and the C-side. Because the geometric coverage of the BCM is quite small, the event 
rate reported by this algorithm during the beam-separation scans was too low to 
perform a reliable calibration. Therefore this 
algorithm will not be considered further in this paper;
\item \bcmEventXOr\ counts the number of events per BC\\
where at least one hit 
above threshold is observed on the C-side, with none observed on the A-side 
within the same 12.5~ns-wide window. Because
converting the event-counting probability measured by this 
method into an instantaneous luminosity involves more complex combinatorics 
than for
the simpler \eventOr\ and \eventAnd\ cases, fully exploiting this algorithm 
requires more extensive studies. These lie beyond the scope of the present 
paper.
\end{itemize}

\subsubsection{LUCID Algorithms}
\label{sec:LUCID-methods}

Four algorithms are currently implemented in the LUMAT card:
\begin{itemize}
\item 
\lucidZeroOr\ counts the number of events per BC where at least one of the two 
detector sides reports no hits within one BCID, 
or where neither side contains any hit in one BCID;
\item
\lucidZeroAnd\ counts the number of events per BC where no hit is found
within one
BCID on either detector side;
\item
\lucidHitOr\ reports the mean number of hits per BC. In this algorithm, hits are 
counted for any event where there is at least one hit in any one of the 16 tubes in 
either detector side in one BCID;
\item
\lucidHitAnd\ reports the mean number of hits per BC, with the additional requirement
that the event contain at least one hit on each of the two detector sides
in one BCID. 
\end{itemize}

\par
The LUCID event-counting algorithms simply subtract the number of empty events 
reported by the zero-counting algorithms above from the total number of bunch 
crossings:
\begin{itemize}
\item \lucidEventAnd\ reports the number of events with at least one hit on each 
detector side ($N_\lucidEventAnd = N_{BC}-N_\lucidZeroOr$);
\item \lucidEventOr\ reports the number of events for which the sum of the hits on 
both detector sides is at least one ($N_\lucidEventOr = N_{BC}-N_\lucidZeroAnd$).
\end{itemize}

\par
Converting measured hit-counting probabilities into instantaneous 
luminosity does not lend itself to analytic models of the type
used for event counting and requires detailed Monte Carlo modeling
that depends on the knowledge of both the detector response and 
the particle spectrum in \pp\ collisions.  This modeling
introduces additional systematic uncertainties and 
to be used reliably requires
more extensive studies that lie beyond the scope of the present paper.

\subsubsection{MBTS Algorithms}
Raw online luminosity information is supplied by the following two CTP scalers:
\begin{itemize}
\item \mbtsone\ counts the number of events per BC where at least one hit above 
threshold is observed on either the A-side or the C-side, or both;
\item \mbtsoneone\ counts the number of events per BC where at least one hit above 
threshold is observed
both on the A- and 
the C-side.
\end{itemize}

\subsubsection{ZDC Algorithms}
Online luminosity information is supplied by dedicated ZDC scalers that count 
pulses produced by constant-fraction discriminators connected to the analog sum of 
ZDC photomultiplier signals on each side separately:
\begin{itemize}
\item \zdcA\ reports the event rate where at least one hit above threshold is 
observed on the A-side, irrespective of whether a hit is simultaneously observed on 
the C-side;
\item \zdcC\ reports the event rate where at least one hit above threshold is 
observed on the C-side, irrespective of whether a hit is simultaneously observed on 
the A-side;
\item \zdcEventAnd\ reports the event rate where at least one hit above threshold 
is observed in coincidence on the A- and C-sides.
This algorithm is still under study and is not considered further in
this paper.
\end{itemize}
The data described here were taken before the ZDC electronic gains and timings
were fully equalized.  
Hence the corresponding visible cross sections for the A- and C-side 
differ by a few per cent.

\subsection{Offline Algorithms}
\label{subsec:offlineAlgs}

Some luminosity algorithms require 
detailed information that is not easily
accessible online. 
These algorithms use data collected with
a minimum bias trigger ({\it e.g.} one of the MBTS triggers) and typically include tighter requirements
to further reduce backgrounds.  Because such analyses can only be performed 
on events that are recorded by the DAQ system, they
are statistically less powerful than the online algorithms.
However, since the MBTS rates per BCID are not available online,
offline algorithms are important for these detectors
for runs where the currents are very different from one bunch to the next.
In addition, these methods
use event selection criteria that are very similar to
final physics analyses.  

\par
Verification that the luminosities obtained from the offline
methods agree well with those obtained from the online techniques 
through the full range of relevant $\mu$  provides an important cross-check
of systematic uncertainties.  
As with the online measurements, 
the LB-averaged instantaneous luminosities are stored in the 
ATLAS conditions database.

\subsubsection{MBTS Timing Algorithm}

The background rate for events passing the  \mbtsoneone\ trigger is a factor of 
about 1000 below the signal.  As a result, online luminosity measurements from 
that trigger can be reliably calculated  without performing a background 
subtraction.  
However, the signal-to-background ratio
is reduced  when the two beams are displaced relative to each other 
(since the signal decreases but the beam-induced backgrounds remain constant).
At the largest beam separations used during the {\it vdM} scans, the 
background rate approaches 10\%\ of the signal.  While these backgrounds
are included in the fit model used to determine the online MBTS
luminosity calibration (see Section~\ref{sec:VdmFitFunction}),
it is useful to cross-check these calibrations by reanalysing the data
with a tighter offline selection.  
The offline time resolution of the MBTS is $\sim 3$~ns and the distance
between the A- and C-sides corresponds to a time difference of 23~ns
for particles moving at the speed of light.
Imposing a requirement
that the difference in time measured for signals from the two sides
be less than 10~ns reduces the background rate in the  \mbtsoneone\ 
triggered events to a negligible level ($<10^{-4}$) even at the largest beam
displacements used in the scans, while maintaining
good signal efficiency.  This algorithm is called \mbtsOFF.
In those instances where different bunches have substantially different
luminosities,  \mbtsOFF\ can be used to properly account for the pile-up dependent 
corrections.

\subsubsection{Liquid Argon Algorithm}
The timing cut used in \mbtsOFF\  is only applicable to coincidence triggers,
where hits are seen both on the A- and C-sides.
It is possible to cross-check the online calibration of the single-sided
\mbtsone\ trigger, where the signal-to-background ratios are lower, 
by imposing timing requirements on a different detector.  
The \LARtiming\ algorithm  uses the liquid argon endcap calorimeters
for this purpose.
Events are required to
pass the \mbtsone\ trigger and to have significant in-time energy deposits
in both EM calorimeter endcaps.  
The analysis considers the energy deposits in the EMEC Inner 
Wheels and the first layer of the FCal,  corresponding to the pseudorapidity 
range $2.5<|\eta|<4.9$. 
Cells are required to have an energy $5\sigma$
above the noise level and to have $E>250$~MeV in the EMEC 
or $E>1200$~MeV in the FCal.
Two cells  are required to pass the selection on each of the A- and C-side.
The time on the A-side (C-side) is then defined as the average time
of all the cells on the A-side (C-side) that pass the above requirements.
The times obtained from the A-side and C-side are then required
to agree to better than  $\pm 5$~ns (the distance
between the A- and C-sides corresponds to a time difference of 30~ns
for particles moving at the speed of light).

\subsubsection{Track-Based Algorithms}
Luminosity measurements have also been performed offline
by counting the rate of events with one or more reconstructed tracks
in the \mbtsone\  sample.  Here, rather than imposing a timing cut,
the sample is selected by requiring that one or more charged particle tracks be
reconstructed in the inner detector.
Two variants of this analysis have been implemented that
differ only in the details of the track selection.

\par
The first method, referred to here as {\it primary-vertex event counting}
(\mbtspvtx ) has larger acceptance. The track selection and vertex reconstruction requirements
are identical to those used for the study of charged particle 
multiplicities at $\sqrt{s}=7$~TeV\cite{bib:CONF046}.
Here, a reconstructed primary vertex is required that is formed from at
least two tracks, each with $p_T>100$~MeV.
Furthermore, the tracks are required
to fulfill the following quality requirements: 
transverse impact parameter  
$|d_0| < 4$~mm with respect to the luminous centroid, errors on the 
transverse and longitudinal impact parameters
$\sigma (d_0)<5$~mm and $\sigma(z_0)<10$~mm, at least 4 hits in 
the SCT, and at least 6 hits in Pixel and SCT. 

\par
The second analysis, referred to here as {\it charged-particle event counting}
(\chpart), is designed to allow the comparison of results from ALICE, ATLAS
and CMS.  It therefore uses fiducial and $p_T$ requirements
that are accessible to all three experiments.
The method counts the rate of events that have
at least one track with transverse momentum $p_T>0.5$~GeV and
pseudorapidity $|\eta| < 0.8$.
The track selection and acceptance corrections  are
identical (with the exception of the $|\eta| < 0.8$ requirement) to 
those in Ref.~\cite{bib:CONF024}.
The main criteria are an \mbtsone\ trigger,  a reconstructed
primary vertex with at least three tracks with $p_T>150$~MeV, and
at least one track with $p_T>500$~MeV,
$|\eta| < 0.8$ and at least 6 SCT hits and one Pixel hit.  Data are
corrected for the trigger efficiency, the efficiency of the vertex
requirement and the tracking efficiency, all of which depend on
\pt\ and $\eta$.

\subsection{Converting Counting Rates to Absolute Luminosity}
\label{subsec:mudependence}
The value of $\mu^{vis}_i$ used to determine the bunch luminosity $\mathcal{L}_i$ 
in BCID $i$ is obtained from the raw number of counts $N_i$ and the number of 
bunch crossings $N_{BC}$, using an algorithm-dependent expression
and assuming that:
\begin{itemize}
\item
the number of \pp-interactions occuring in any bunch crossing obeys a Poisson 
distribution. This assumption drives the combinatorial formalism presented in 
Sections~\ref{subsubsec:muEvOr} and ~\ref{subsubsec:muEvAND} below;
\item
the efficiency to detect a single inelastic \pp\ interaction is constant, in the 
sense that it does not change when several interactions occur in the same bunch 
crossing. This is tantamount to assuming that the efficiency $\varepsilon_n$
for detecting one event 
associated with  $n$ interactions occuring in the same crossing
is given by
\begin{equation}
\varepsilon_n = 1 - (1-\varepsilon_1 )^n
\label{eq:eff_def}
\end{equation}
where $\varepsilon_1$ is the detection efficiency corresponding to a single 
inelastic interaction in a bunch crossing (the same definition applies to the 
efficiencies $\varepsilon^{OR}$, $\varepsilon^A$, $\varepsilon^C$ and 
$\varepsilon^{AND}$ defined below). This assumption will be validated in 
Section~\ref{subsubsec:EvPileupck}. 
\end{itemize}
The bunch luminosity is then given directly and without additional 
assumptions by
\begin{equation}
\mathcal{L}_i = 
\frac{{\mu^{vis}_i}f_r}{{\sigma _{vis} }}
\end{equation}
using the value of $\sigma_{vis}$ measured during beam-separation scans for the 
algorithm considered. 
However, providing a value for
$\mu \equiv \muvis / \varepsilon = \muvis \sigma_{inel}/\sigma_{vis}$ 
requires an assumption on the 
as yet unmeasured total inelastic cross section
at $\sqrt{s}=7$~TeV\footnote{
ATLAS uses the \pythia\ value of 71.5~mb.}.

\subsubsection{Inclusive-OR Algorithms}
\label{subsubsec:muEvOr}

In the \eventOr\ case, the logic is straightforward. Since the Poisson probability  
for observing zero events in a given bunch crossing is 
$P_0(\mu^{vis}) = e^{-\mu^{vis}} = e^{-\mu \varepsilon^{OR}}$, the probability of 
observing at least one event is 
\begin{equation}
\begin{array}{lcl}
P_\tinyeventOr(\mu^{vis}) & = & \frac{N_{OR}}{N_{BC}} \\
    & = & 1-P_0(\mu^{vis})  \\
    & = & 1-e^{-\mu^{vis}} \\
\end{array}
\label{eq:PForEventOr}
\end{equation}
Here the raw event count $N_{OR}$ is the number of bunch crossings, during a given 
time, in which at least one \pp\ interaction satisfies the event-selection criteria 
of the OR algorithm under consideration, and $N_{BC}$ is the total number of bunch 
crossings during the same interval. Equation~\ref{eq:PForEventOr} reduces to the 
intuitive result $P_\tinyeventOr(\mu^{vis})\approx \muvis$ when $\mu_{vis} << 1$.
Solving for $\mu^{vis}$ in terms of the event-counting rate yields:
\begin{equation}
\begin{array}{ll}
\mu^{vis} = - \ln \left( 1- \frac{N_{OR}}{N_{BC}} \right )\\
\end{array}
\label{eq:muForEventOr}
\end{equation}

\subsubsection{Coincidence Algorithms}
\label{subsubsec:muEvAND}

For the \eventAnd\ case, the relationship between \muvis\ and $N$ is
more complicated. Instead of depending on a single efficiency, the event-counting 
probability must be written in terms of $\varepsilon^A$, $\varepsilon^C$ and 
$\varepsilon^{AND}$, the efficiencies for observing an event with, respectively, at 
least one hit on the A-side, at least one hit on the C-side and at least one hit on 
both sides simultaneously. These efficiencies are related to the \eventOr\ 
efficiency by 
$\varepsilon^{OR}= \varepsilon^A + \varepsilon^C - \varepsilon^{AND}$.

\par
The probability $P_\tinyeventAnd(\mu)$ of there being at least one hit on both sides is 
one minus the probability $P_0^{\zeroOr}$ of there being no hit on at least one 
side. The latter, in turn, equals the probability that there be no hit on at least 
side A ($P_{0A} = e^{-\mu\varepsilon^A}$), plus the probability that there be no 
hit on at least side C ($P_{0C} = e^{-\mu\varepsilon^C}$), minus the probability 
that there be no hit on either side ($P_0 = e^{-\mu\varepsilon^{OR}}$):
\begin{equation}
\begin{array}{lll}
P_\tinyeventAnd(\mu) 
            & = & \frac{N_{AND}}{N_{BC}} \\
            & = & 1 - P_0^{\zeroOr}(\mu) \\ 
            & = & 1- (e^{-\mu\varepsilon^{A}}+ e^{-\mu\varepsilon^{C}}
                      - e^{-\mu\varepsilon^{OR}} ) \\
            & = & 1- (e^{-\mu\varepsilon^{A}}+ e^{-\mu\varepsilon^{C}}
      -e^{-\mu (\varepsilon^{A} + \varepsilon^{C} - \varepsilon^{AND})} ) \\
\end{array}
\label{eq:eventAndProb1}
\end{equation}
This equation cannot be inverted analytically. The most appropriate functional form 
depends on the values of  $\varepsilon^A$, $\varepsilon^C$ and $\varepsilon^{AND}$.  

\par
For cases such as \lucidEventAnd\ and \bcmEventAnd, the above equation can be 
simplified using the fact that $\varepsilon^{AND} << \varepsilon^{A,C}$, and 
assuming that $\varepsilon^A \approx \varepsilon^C$. The efficiencies 
$\varepsilon^{AND}$ and $\varepsilon^{OR}$ are {\it defined} by, respectively, 
$\varepsilon^{AND} \equiv \sigma_{vis}^{AND}/\sigma_{inel}$ and
$\varepsilon^{OR} \equiv \sigma_{vis}^{OR}/\sigma_{inel}$; the average number of 
visible inelastic interactions per BC is computed as 
$\muvis \equiv \varepsilon^{AND} \mu$. Equation~\ref{eq:eventAndProb1} then becomes
\begin{equation}
\begin{array}{lll}
\frac{N_{AND}}{N_{BC}}
    & = & 1- 2e^{-\mu(\varepsilon^{AND}+\varepsilon^{OR})/2}
          + e^{-\mu\varepsilon^{OR}}                               \\
    & = & 1- 2e^{-(1+ {\sigma_{vis}^{OR}}/{\sigma_{vis}^{AND}})\muvis/2}
           + e^{-({\sigma_{vis}^{OR}}/{\sigma_{vis}^{AND}})\muvis} \\
\end{array}
\label{eq:eventAndProb2}
\end{equation}
The value of $\mu_{vis}$ is then obtained by solving
Equation~\ref{eq:eventAndProb2} numerically
using the values of $\sigma_{vis}^{OR}$ and $\sigma_{vis}^{AND}$   
extracted from beam separation scans.
The validity of this technique will be quantified in 
Section~\ref{sec:Cnstcy}.
\par
If the efficiency is high and 
$\varepsilon^{AND} \approx \varepsilon^{A}\approx \varepsilon^{C}$, as is the case 
for  \mbtsoneone, Equation~\ref{eq:eventAndProb1} can be approximated by
\begin{equation}
\begin{array}{ll}
                \mu^{vis} \approx 
                - \ln \left( 1- \frac{N_{AND}}{N_{BC}} \right )\\
\end{array}
\label{eq:mbtsApprox}
\end{equation}

\par
The $\mu$-dependence of the probability function $P_\tinyeventAnd$ is 
controlled by the relative magnitudes of $\varepsilon^A$, $\varepsilon^C$ and 
$\varepsilon^{AND}$ 
(or of the corresponding measured visible cross sections). This is in contrast to 
the \eventOr\ case, where the efficiency $\varepsilon_{OR}$ factors out of 
Equation~\ref{eq:muForEventOr}.

\subsubsection{Pile--up-related Instrumental Effects}
\label{subsubsec:EvPileupck}

\begin{figure}  
\begin{center}   
\epsfig{file=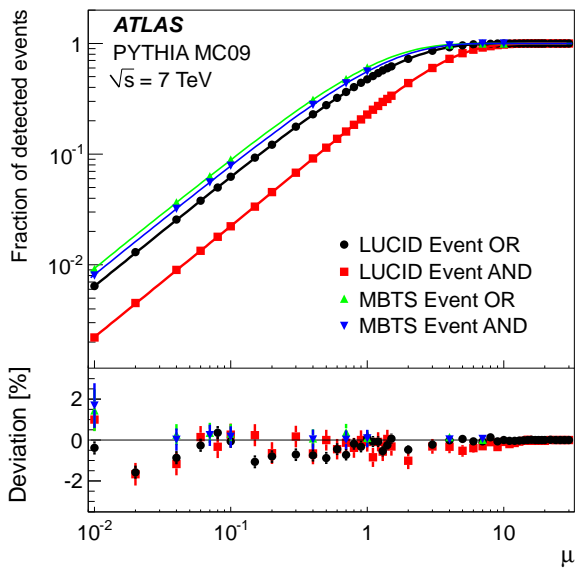,width=0.45\textwidth}
\end{center}
\caption{Fraction of bunch crossings containing a detected event
for LUCID and MBTS 
algorithms as a function of $\mu$, the true average number of inelastic \pp\
interactions per BC. The plotted points are the result of a Monte Carlo study performed 
using the \pythia\ event generator together with a GEANT4 simulation of the ATLAS 
detector response. The curves reflect the combinatorial formalism of 
Sections~\ref{subsubsec:muEvOr} and~\ref{subsubsec:muEvAND}, using as input only 
the visible cross sections extracted from that same simulation. 
The bottom inset shows the difference between the full simulation 
and the parameterization.}
\label{fig:MCvsParameterizationFnMu}
\end{figure}

The $\mu$-dependence of the probability functions $P_\tinyeventOr$ and\\
$P_\tinyeventAnd$ is 
displayed in Fig.~\ref{fig:MCvsParameterizationFnMu}. All algorithms {\it 
saturate} at high $\mu$, reflecting the fact that as the pile-up increases, the 
probability of observing at least one event per bunch crossing approaches one. Any 
event-counting luminosity algorithm will therefore 
lose precision, and ultimately become unusable, as the LHC luminosity per bunch 
increases far beyond present levels. The tolerable pile-up level is detector- and 
algorithm-dependent: the higher the efficiency
($\varepsilon^{OR}_{MBTS} > \varepsilon^{AND}_{MBTS} >
  \varepsilon^{OR}_{LUCID} > \varepsilon^{AND}_{LUCID}$),
the earlier the onset of this saturation.  

\par
The accuracy of the event-counting formalism can be verified using simulated data. 
Figure~\ref{fig:MCvsParameterizationFnMu} (bottom) shows that the parameterizations 
of Sections~\ref{subsubsec:muEvOr} and \ref{subsubsec:muEvAND} deviate from the full 
simulation 
by $\pm 2$\%\ at most: possible instrumental  effects not accounted for by the 
combinatorial formalism are predicted to have 
negligible impact for the bunch luminosities achieved in the 2010 LHC run ($0<\mu<5$).

It should be stressed, however, that the agreement between the Poisson formalism 
and the full simulation depends critically on the validity of the assumption, 
summarized by Equation~\ref{eq:eff_def}, that the efficiency for detecting an 
inelastic \pp\ interaction is independent of the number of interactions that occur 
in each crossing. This requires, for instance, that the threshold for registering a 
hit in a phototube (nominally 15 photoelectrons for LUCID) be low enough compared 
to the average single-particle response. This condition is satisfied by the 
simulation shown in Fig.~\ref{fig:MCvsParameterizationFnMu}. Repeating this 
simulation with the LUCID threshold raised to 50 photoelectrons yields systematic 
discrepancies as large as 7\%\ between the computed and simulated probability 
functions for the LUCID  \eventAnd\ algorithm. When the threshold is too high, a 
particle from a single \pp\ interaction occasionally fails to fire the 
discriminator. 
However, if two such particles from different \pp\ interactions in 
the same bunch crossing traverse the same tube, they may produce enough light to 
register a hit.
This effect is called migration.

\section{Absolute Calibration Using Beam-Separation Scans}
\label{sec:Vdm}

The primary calibration of all luminosity algorithms is derived from data collected 
during van der Meer scans. The principle (Section~\ref{sec:method}) is to measure 
simultaneously the collision rate at zero beam separation and the corresponding absolute
luminosity inferred from the charge of the 
colliding proton bunches and from the horizontal and vertical convolved beam 
sizes~\cite{bib:Kozanecki:2008zz}. Three sets of beam scans have been carried out in ATLAS, 
as detailed in Section~\ref{sec:dataSets}. These were performed in both the 
horizontal and the vertical directions in order to reconstruct the transverse convolved 
beam profile. During each scan, the collision rates measured by the  luminosity 
detectors were recorded while the beams were moved stepwise with respect to each 
other in the transverse plane.

\subsection{Absolute Luminosity from Beam Parameters}
\label{sec:method}

In terms of colliding-beam parameters, the luminosity $\mathcal L$ is defined (for beams that collide with zero crossing angle) as
\begin{equation}
{\mathcal L} = n_b f_r n_1 n_2 \int {\hat{\rho} _1 (x,y)} \hat{\rho} _2
(x,y)dxdy\label{lumi}
\end{equation}
where $n_b$ is the number of colliding bunches, $f_r$ is the machine revolution 
frequency ($11245.5\;Hz$ for LHC), $n_{1(2)}$ is the number of particles per 
bunch in beam 1 (2) and $\hat{\rho}_{1(2)}(x,y)$ is the normalized particle density 
in the transverse ($x$-$y$) plane of beam 1 (2) at the IP. Under the general 
assumption that there is no correlation between $x$ and $y$, {\it i.e.} that 
transverse coupling is negligible\footnote{Combining the observed vertical to horizontal 
emittance ratios with the measured LHC lattice functions indicates that the 
luminosity loss caused by the residual tilt of the two beams was less than 0.25\%.}
at the IP, the particle densities can be 
factorized ($\hat{\rho}(x,y)=\rho(x)\rho(y)$) and
 Equation~\ref{lumi} rewritten as
\begin{equation}
{\mathcal L} = n_b f_r n_1 n_2 
               ~\Omega _x (\rho _1 (x),\rho _2(x))
               ~\Omega _y (\rho _1 (y),\rho _2 (y))
\label{lumi1}
\end{equation}
where \[ \Omega _x (\rho _1,\rho _2 ) = \int {\rho _1 (x)}\rho _2 (x)dx\] 
is the beam overlap integral in the $x$ direction (with an analogous
definition in the $y$ direction). In the
method proposed by van der Meer~\cite{bib:vdm} the overlap integral
(for example in the $x$ direction) can be calculated as:
\begin{equation}
\Omega _x (\rho _1,\rho _2) = \frac{{R_x (0)}}{{\int {R_x
(\delta)d\delta} }}\label{vdm}
\end{equation}
where $R_x(\delta)$ is the luminosity (or equivalently $\mu^{vis}$) -- at this stage in 
arbitrary units -- measured during a horizontal scan at the time the two beams are 
separated by the distance $\delta$ and $\delta=0$ represents the 
case of zero beam
separation. 
$\Sigma_x$ is defined by the equation:
\begin{equation}
\Sigma _x  = \frac{1}{{\sqrt {2\pi } }}\frac{{\int {R_x (\delta)d\delta}
}}{{R_x (0)}}
\label{caps}
\end{equation}
In the case where the luminosity curve $R_x(\delta)$ is Gaussian, 
$\Sigma _x $ coincides 
with the standard deviation of that distribution. By using the last two equations, 
Equation~\ref{lumi1} can be rewritten as
\begin{equation}
{\mathcal L} = \frac{{n_b f_r n_1 n_2 }}{{2\pi \Sigma _x \Sigma _y
}}\label{lumifin}
\end{equation}
which is a general formula to extract luminosity from machine parameters by 
performing a beam separation scan. Equation~\ref{lumifin} is quite 
general; $\Sigma_x$ and $\Sigma_y$ only depend  on the area under the luminosity curve.

\subsection{Luminosity-Scan Data Sets}
\label{sec:dataSets}
Three van der Meer scans have been performed at the ATLAS
interaction point (Table~\ref{tab:vdmScan}).
The procedure~\cite{bib:helmut,bib:simon-ipac-2010} ran as follows.
After centring the beams on each other at the IP in both the horizontal 
and the 
vertical plane using mini-scans, a full  luminosity-calibration 
scan was carried out in the horizontal plane, 
spanning a range of $\pm 6 \sigma_b$ in horizontal beam-separation
(where $\sigma_b$ is the nominal transverse size of either beam at the IP). 
A full luminosity-calibration scan was then carried
out in the vertical plane, again spanning a range of $\pm 6 \sigma_b$ in 
relative beam separation. 

The mini-scans used to first centre the beams on each other in the
transverse plane were done by  activating closed orbit 
bumps\footnote{A closed orbit bump is a local distortion of the beam orbit
that is implemented using pairs of steering dipoles located on either
side of the affected region. In this particular case, these bumps are
tuned to translate either beam parallel to itself at the IP, in either the 
horizontal or the vertical direction.} 
around the IP that vary the IP positions of both beams by $\pm 1\sigma_b$
in opposite directions, either horizontally or vertically. 
The relative positions of the two
beams were then adjusted, in each plane, to achieve (at that time)
optimum transverse overlap.

The full horizontal and vertical scans followed an identical 
procedure, where the same orbit bumps were used to displace the two beams in 
opposite directions by $\pm 3 \sigma_b$, resulting in a total 
variation of $\pm 6 \sigma_b$ in relative displacement at the IP. 
In Scan~I, the horizontal scan started at zero nominal separation, moved to the
maximum  separation in the negative direction, stepped back to 
zero and on to the maximum positive separation, and finally returned to 
the original settings of the closed-orbit bumps (zero nominal separation). 
The same procedure was followed for the vertical scan.
In Scan~II and~III, after collision optimization with the transverse 
mini-scans, a full horizontal scan was  taken from 
negative to positive nominal separation, followed by a hysteresis cycle 
where the 
horizontal nominal separation was run to $-6 \sigma_b$, then
0 then $+6 \sigma_b$, and finally followed by a full  horizontal scan
in the opposite direction to check for potential hysteresis effects. 
The same procedure was then repeated in the vertical direction.

For each scan, at each of 27 steps in
relative displacement, the beams were left in a quiescent state for $\sim
30$\,seconds. During this time the (relative) luminosities 
measured by all active 
luminosity monitors were recorded 
as a function of time in a dedicated online-data stream, together 
with the value of 
the nominal separation, the beam currents and  other relevant accelerator 
parameters transmitted to ATLAS by the
accelerator control system.  In addition, the full data acquisition 
system was operational throughout the scan, using the standard 
trigger menu, and triggered events were recorded as part of the normal
data collection.  
\begin{center}
\begin{table*}
\begin{center}
\begin{tabular}[h]{|l|c|c|}
\hline
               & vdM Scan~I                & vdM Scan~II, III\\
               & (April 26, 2010)            & (May 9, 2010)     \\
\hline
LHC Fill Number & 1059 & 1089 \\
\hline
Scan Directions & 1 horizontal scan           & 2 horizontal scans      \\
               & followed by 1 vertical scan & followed by 2 vertical scans \\
\hline
Total Scan Steps per Plane &  27              &  54 (27+27) \\
                           & $(\pm 6 \sigma_b)$ & $(\pm 6 \sigma_b) $
\\\hline
  Scan Duration per Step   & 30~sec            & 30~sec \\
\hline
 Number of bunches colliding in ATLAS      &  1               & 1 \\
\hline
 Total number of bunches per beam          &  2               & 2 \\
\hline
  Number of protons per bunch
                       & $\sim 0.1\cdot 10^{11}$ & $\sim 0.2\cdot10^{11}$ \\
\hline
    $\beta^*$ (m)                &   $\sim 2$   &  $\sim 2$  \\
\hline
    $\sigma_b$ ($\mu$m) [assuming nominal emittances]          &  $\sim 45$   & $\sim 45 $ \\
\hline
    Crossing angle ($\mu$rad)    &  0    & 0   \\
\hline
    Typical luminosity/bunch ($\mu{\rm b}^{-1}/s)$ & $  4.5\cdot10^{-3}$ 
                                                   & $  1.8\cdot10^{-2}$ \\
\hline
    $\mu$ (interactions/crossing) & 0.03 & 0.11 \\
\hline
\end{tabular}
\end{center}
\caption{Summary of the main characteristics of the three beam
scans performed at the ATLAS interaction point. The values of
luminosity/bunch and $\mu$
are given for zero beam separation.}
\label{tab:vdmScan}
\end{table*}
\end{center}

\subsection{Parametrization and Analysis of the Beam Scan Data}
\label{sec:VdmFitFunction}

Data from all three scans have been analyzed both from the dedicated
online-data stream and from 
the standard ATLAS data stream.  
Analyses using the standard data stream suffer from reduced
statistical precision relative to the dedicated stream, but allow for
important cross-checks both of the background rates
and of the size and position
of the luminous region.  In addition, because this stream contains full 
events, these data can be used to measure the visible cross section 
corresponding to standard analysis selections that require, for example,
timing cuts in the MBTS or the liquid argon Calorimeter or the presence of
a reconstructed primary vertex.  Measurements performed using these
two streams provide a consistent interpretation of the data within the
relevant statistical and systematic uncertainties.

In all cases, the analyses fit the relative variation of the bunch luminosity as a 
function of the beam separation to extract $\Sigma_x$ and $\Sigma_y$ 
(Equation~\ref{caps}).  These results are then combined with the measured bunch 
currents to determine the absolute luminosity using Equation~\ref{lumifin}.  
Although the pile-up effects remained relatively weak during these scans, the raw 
rates ($P_\eventOr$, $P_\eventAnd$,...) are converted \footnote{For the 
coincidence algorithms, the procedure is iterative because it requires the {\it a 
priori} knowledge of $\sigma_{vis}$. Monte Carlo estimates were used as the 
starting point.} 
into a mean number of interactions per crossing $\muvis$ as described in 
Section~\ref{subsec:mudependence}.
In addition, to remove sensitivity to the slow decay of the beam currents over the 
duration of the scan, the data are analyzed as {\em specific rates}, obtained by 
dividing the measured average interaction rate per BC by the product of the 
bunch currents measured at that scan point:
\begin{equation}
R_{sp} = \frac{(n_1 n_2)_{MAX}}{(n_1 n_2)_{meas}} R_{meas}
\end{equation}
Here $(n_1 n_2)_{meas}$ is the product of the numbers of protons in the two 
colliding bunches during the measurement, $(n_1 n_2)_{MAX}$ is its maximum value 
during the scans, and $R_{meas}$ is the value of  $\muvis$ 
at the current scan point.
\par
Beam currents are measured using two complementary LHC 
systems\cite{bib:currentMease}. The fast bunch-current transformers (FBCT) are 
AC-coupled, high-bandwidth devices which use gated electronics to perform continuous 
measurements of individual bunch charges for each beam. The Direct-Current Current 
Transformers (DCCT) measure the total circulating intensity in each of the two beams 
irrespective of their underlying time structure. The DCCT's have intrinsically 
better accuracy, but require averaging over hundreds of seconds to achieve the needed 
precision. The relative (bunch-to-bunch) currents are based on the FBCT measurement. 
The absolute scale of the bunch intensities $n_1$ and $n_2$ is determined by 
rescaling the total circulating charge measured by the FBCTs to the more accurate 
DCCT measurements. Detailed discussions of the performance and calibration of these 
systems are presented in Ref.~\cite{bib:Gras}.
\par
\begin{figure*}[p]
\begin{center}
\subfigure[]{
\epsfig{file=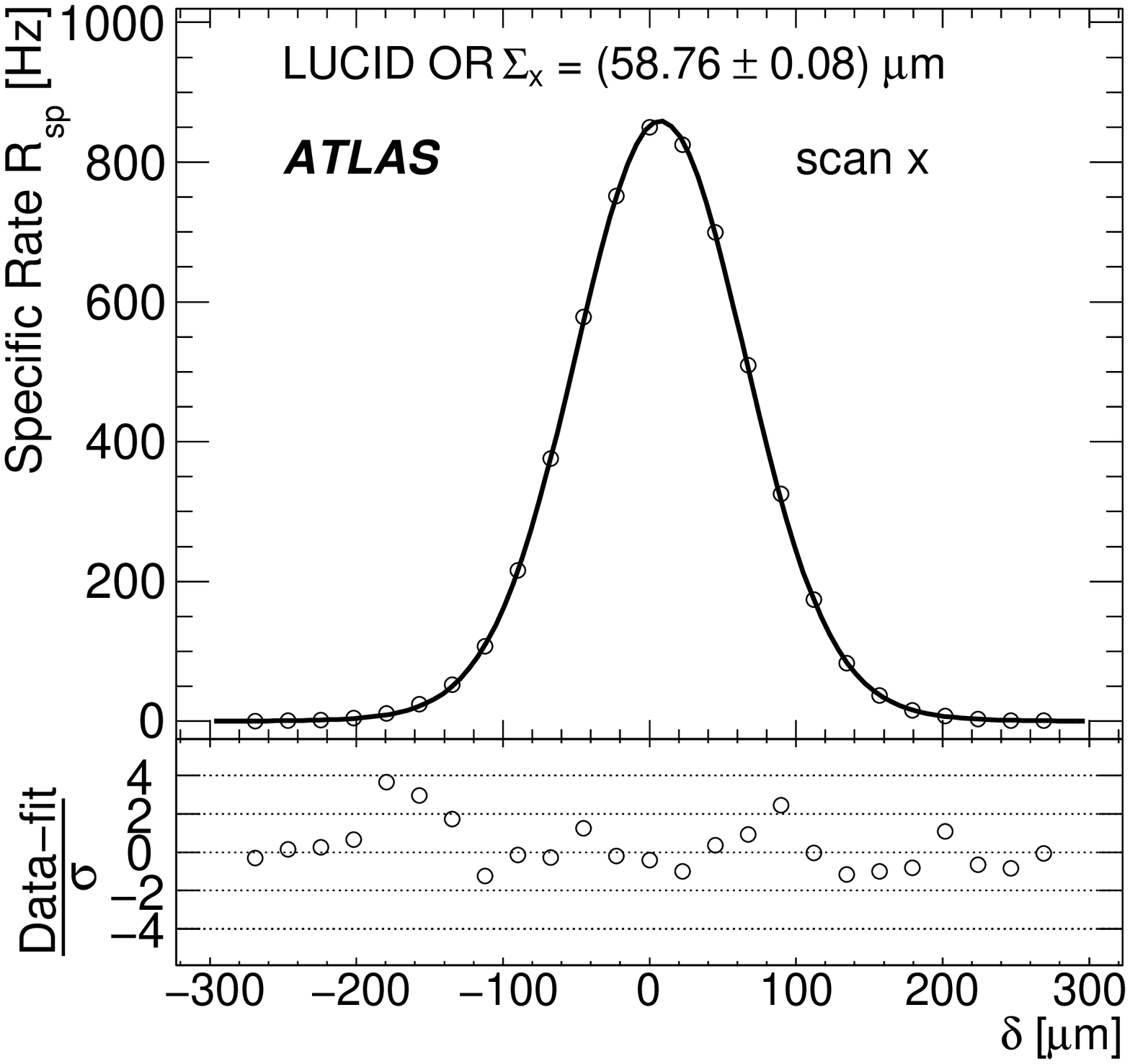,width=0.4\textwidth}
\hfil
\epsfig{file=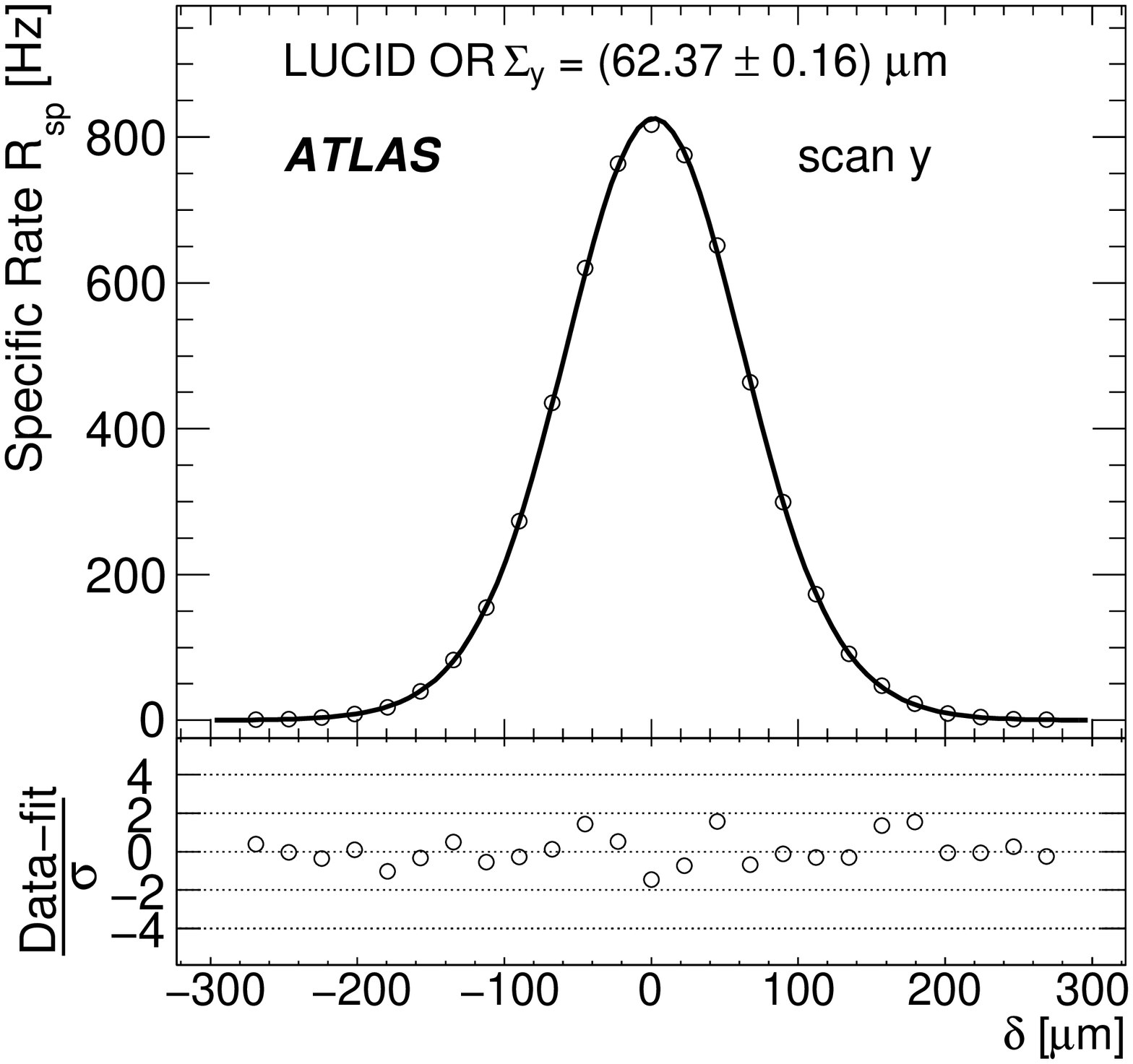,width=0.4\textwidth}
}
\vfil
\subfigure[]{
\epsfig{file=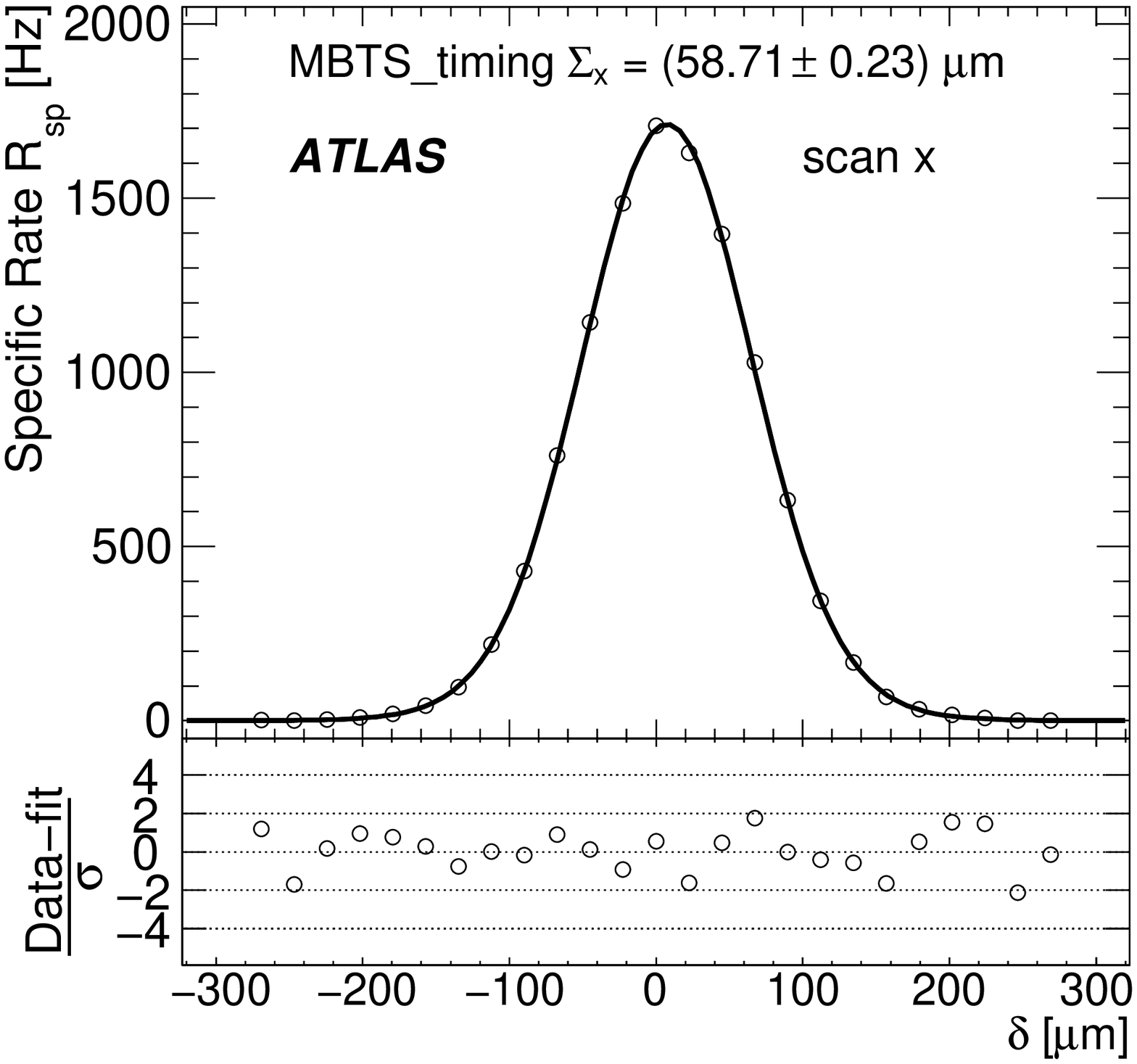,width=0.4\textwidth}
\hfil
\epsfig{file=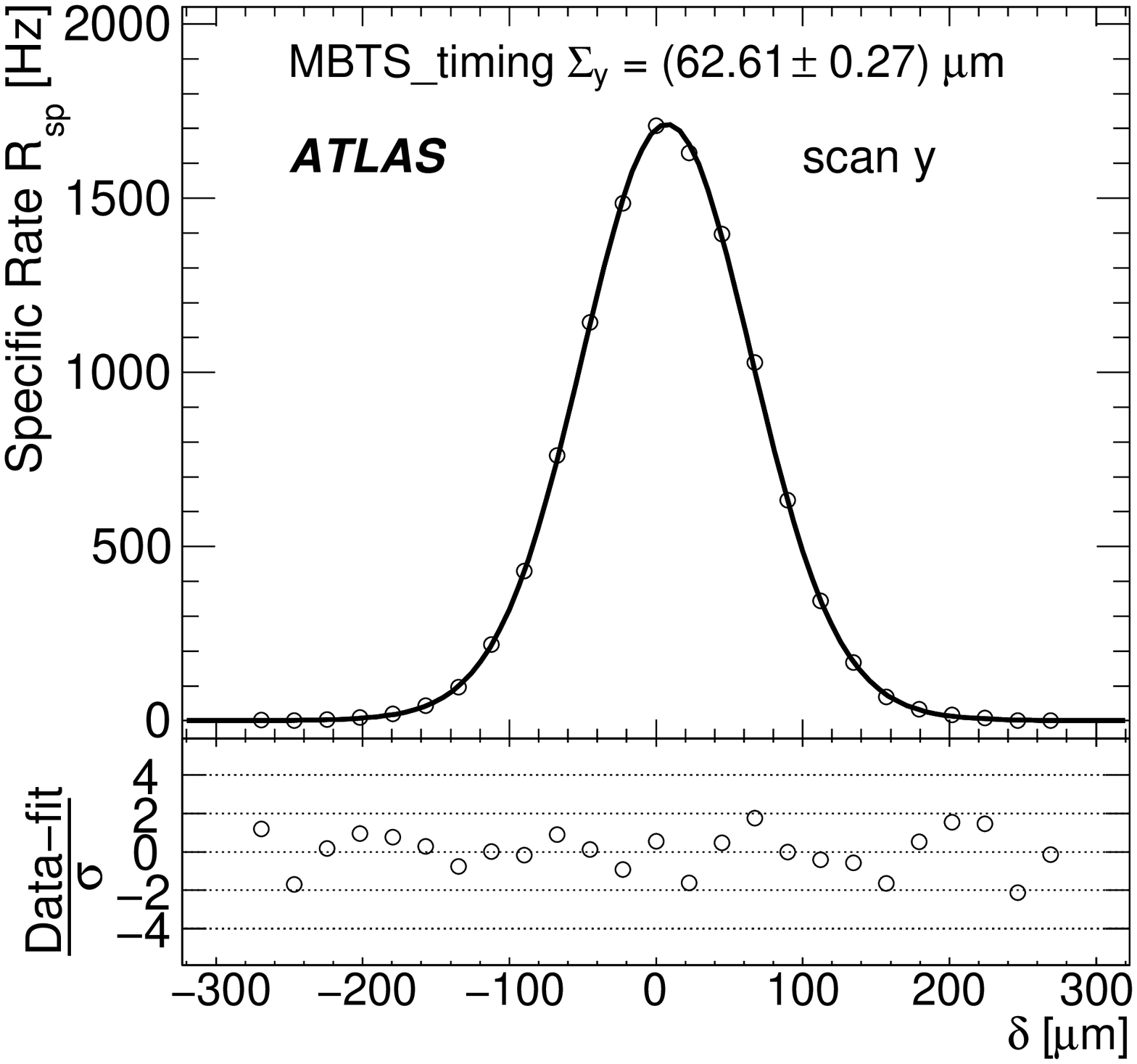,width=0.4\textwidth}
}
\vfil
\subfigure[]{
\epsfig{file=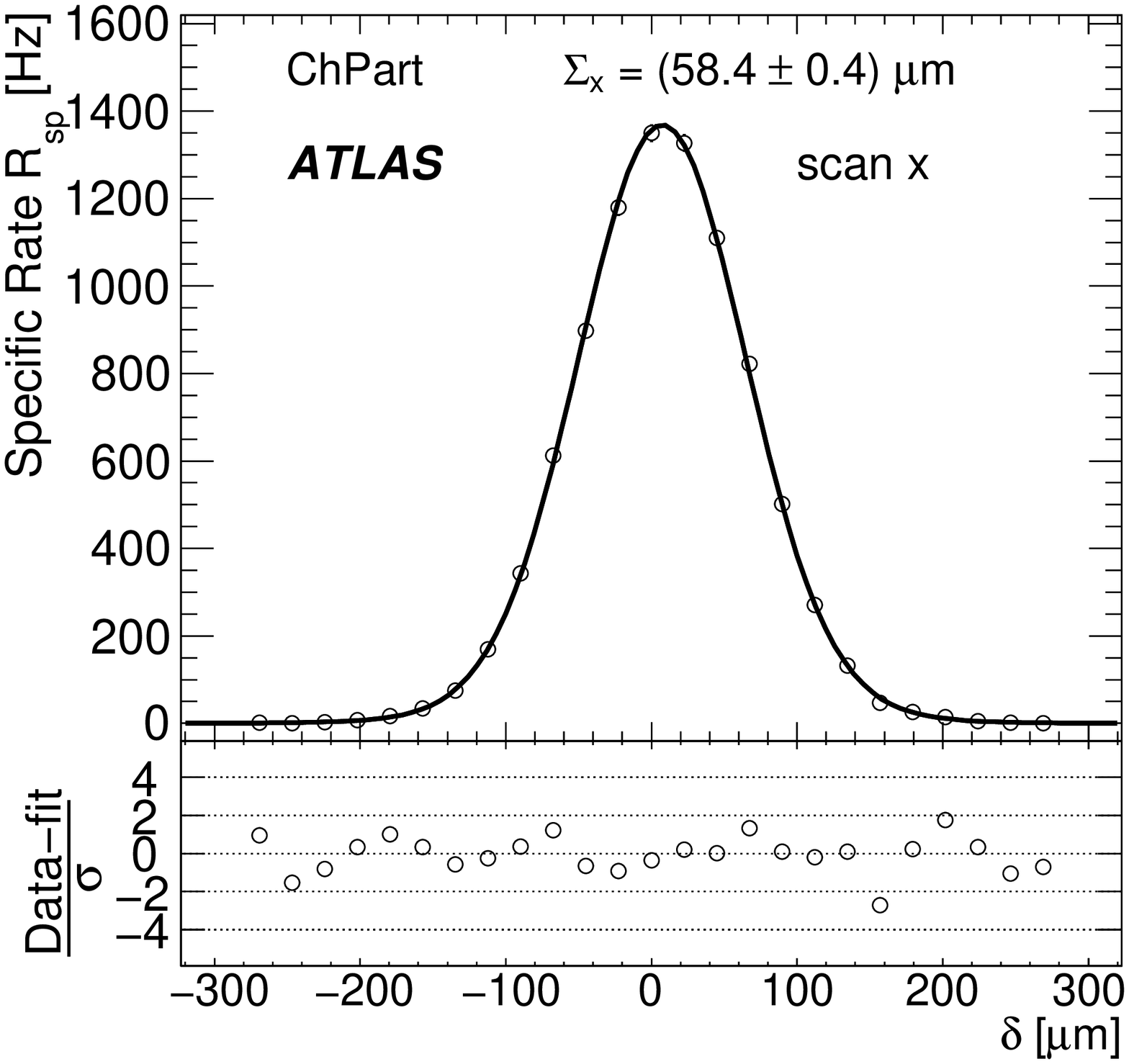,width=0.4\textwidth}
\hfil
\epsfig{file=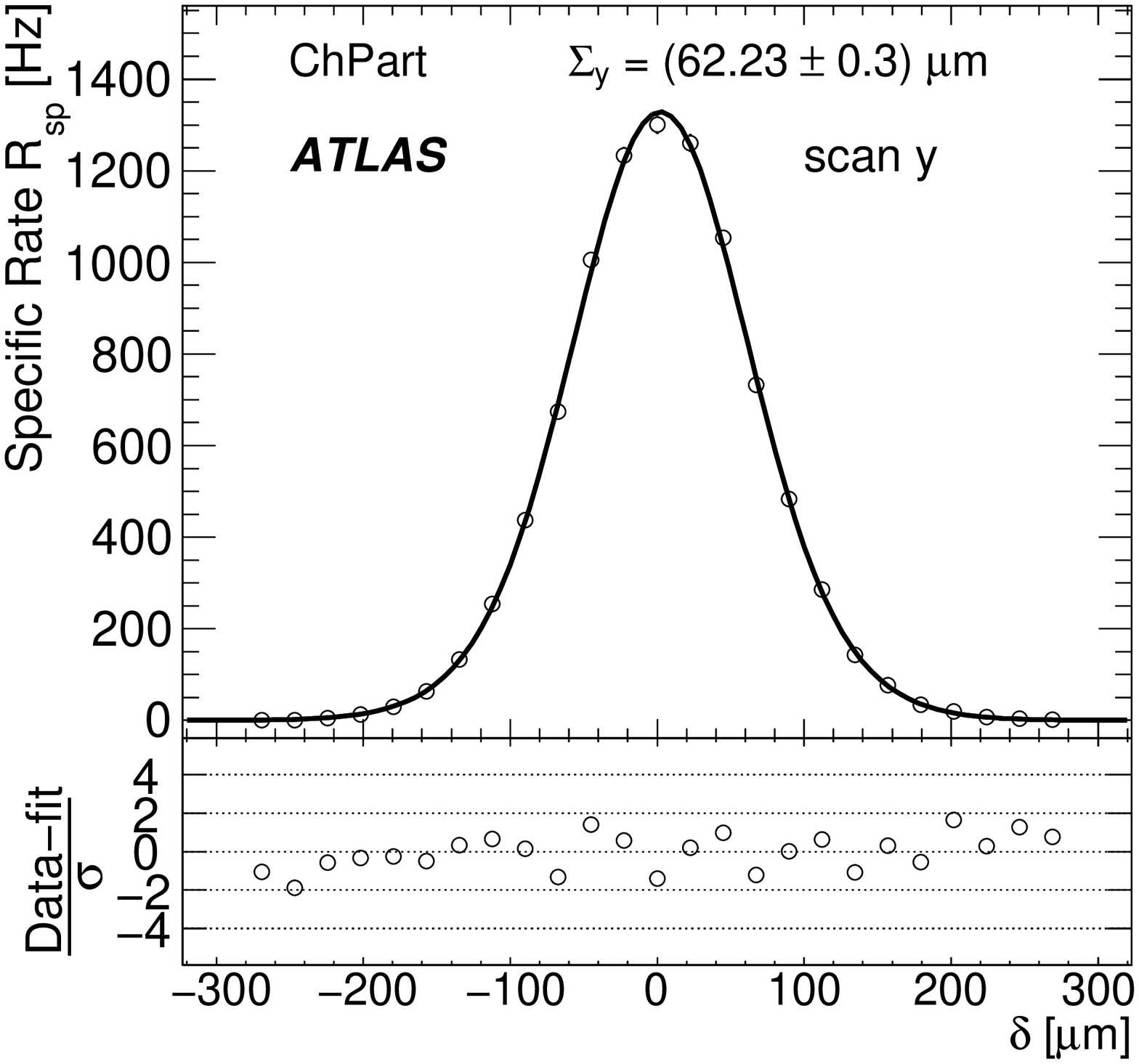,width=0.4\textwidth}
}
\end{center}
\caption{Results of fits to the second luminosity scan in the x (left) 
and y (right) direction for the (a) \lucidEventOr, (b) \mbtsOFF, 
and (c) \chpart\ algorithms. 
The panels at the bottom of each graph show the difference 
of the measured rates from the value predicted by the fit,
normalized to the statistical uncertainty on the data ($\sigma$).
}
\label{fig:vdmFitsI}
\end{figure*}

Fits to the relative luminosity require a choice of parametrization
of the shape of the scan curve.  For all detectors and
algorithms, fits using  a single Gaussian or  
a single Gaussian with a flat background yield unacceptable $\chi^2$
distributions.  In all cases, fits to a double Gaussian (with a common
mean) plus a
flat background result in a $\chi^2$ per degree of freedom close to one.
In general, the background rates are consistent with
zero for algorithms requiring a coincidence between sides, while
small but statistically 
significant backgrounds are observed for algorithms requiring only
a single side.  These backgrounds are reduced to
less than $0.3\% $ of the luminosity at zero beam separation
by using data from the paired bunches only.
Offline analyses that require timing or a primary vertex, in addition to 
being restricted to paired bunches, have very low background.
The residual background is subtracted using the rate measured in 
unpaired bunches; no background
term is therefore needed in the fit function for the offline case. 
Examples of such fits are shown
in Fig.~\ref{fig:vdmFitsI}.

For these fits the specific rate is described by a double Gaussian:
\begin{eqnarray}\nonumber
R_x(\delta) & = & R_x(x-x_0)  \\
& = &  \int \frac{ R_x(\delta) d\delta} {\sqrt{2\pi}} \left [
\frac{f_i e^{-\frac{(x-x_0)^2}{2\sigma_i^2}}}{\sigma_i}
+ \frac{(1-f_i)e^{-\frac{(x-x_0)^2}{2\sigma_j^2}}}{\sigma_j}  \right ]
\end{eqnarray}
Here $\sigma_i$ and $\sigma_j$ are the widths of first and
second Gaussians respectively, 
$f_i$ is the fraction of the rate in the first Gaussian and
$x_0$ is introduced to allow for the possibility that the beams
are not perfectly centred at the time of the scan.
The value of $\Sigma_x$ in Equation~\ref{lumifin} is calculated as
\begin{equation}
\frac{1}{\Sigma_x} = \left [
\frac{f_i}{\sigma_i} + \frac{1-f_i}{\sigma_j}  \right ]
\end{equation}

\subsection{Fit Results}
\label{sec:VdmResults}

\begin{figure*}[p]
\subfigure[]{
\epsfig{file=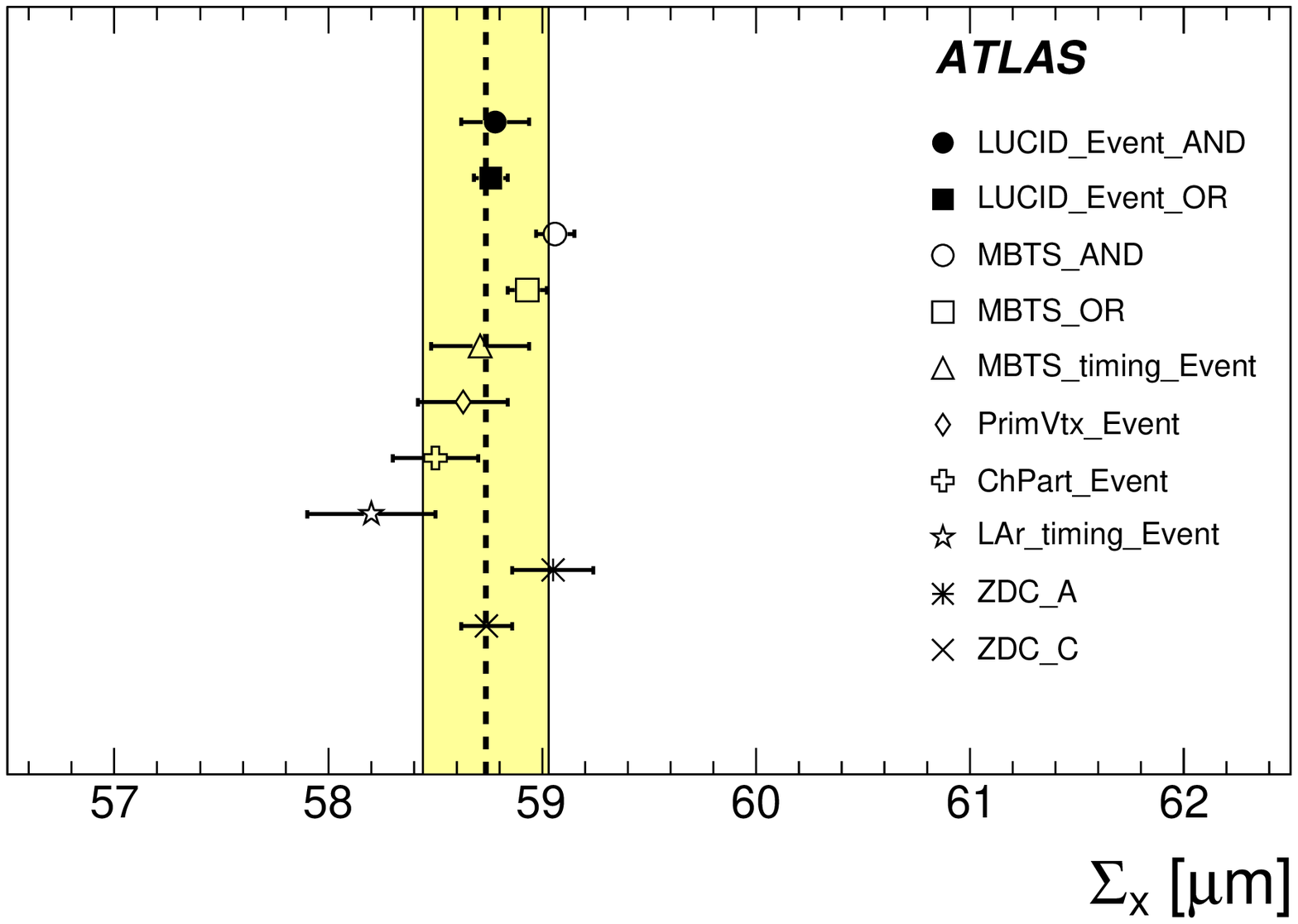,width=0.48\textwidth}
}
\subfigure[]{
\epsfig{file=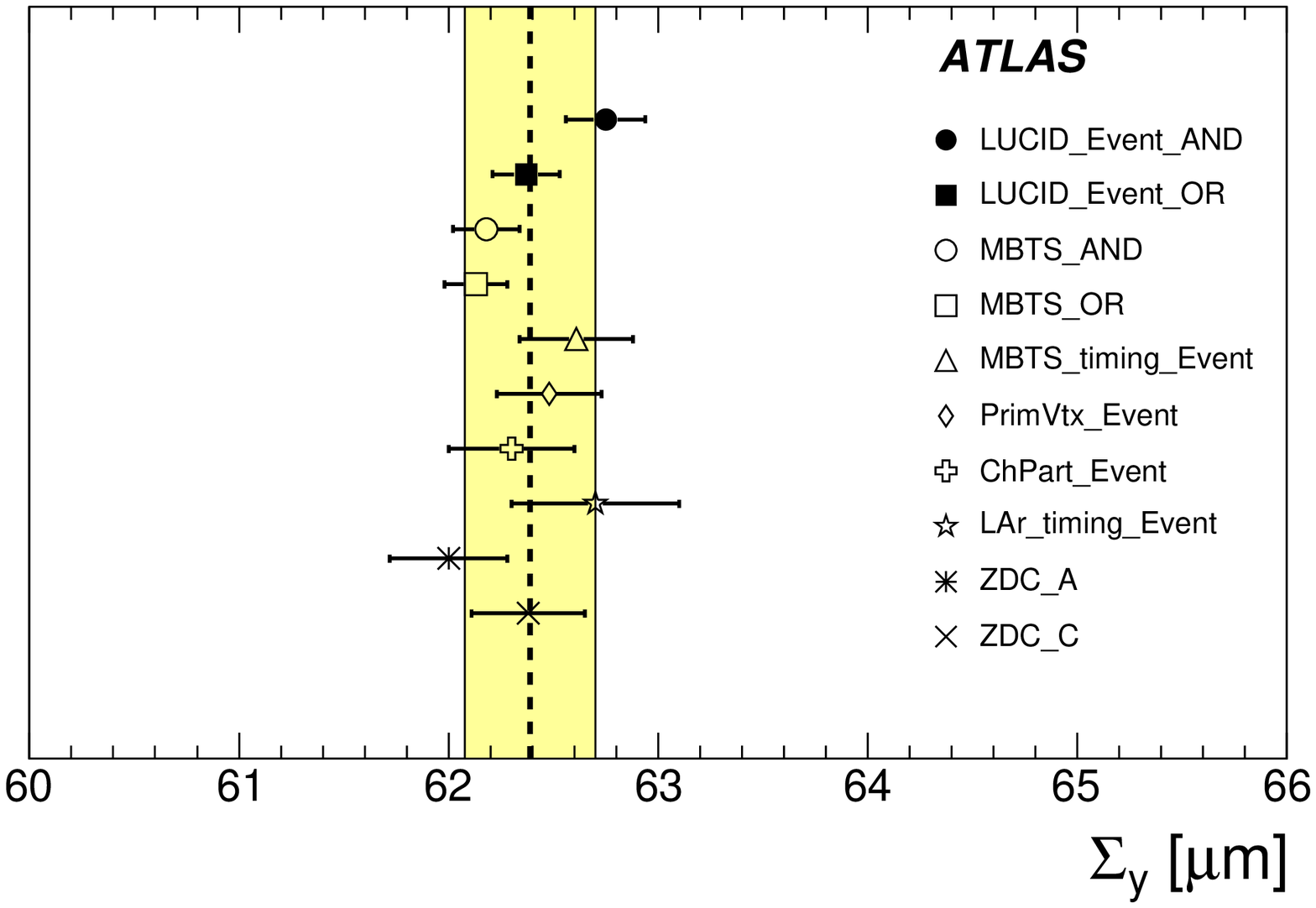,width=0.48\textwidth}
}
\subfigure[]{
\epsfig{file=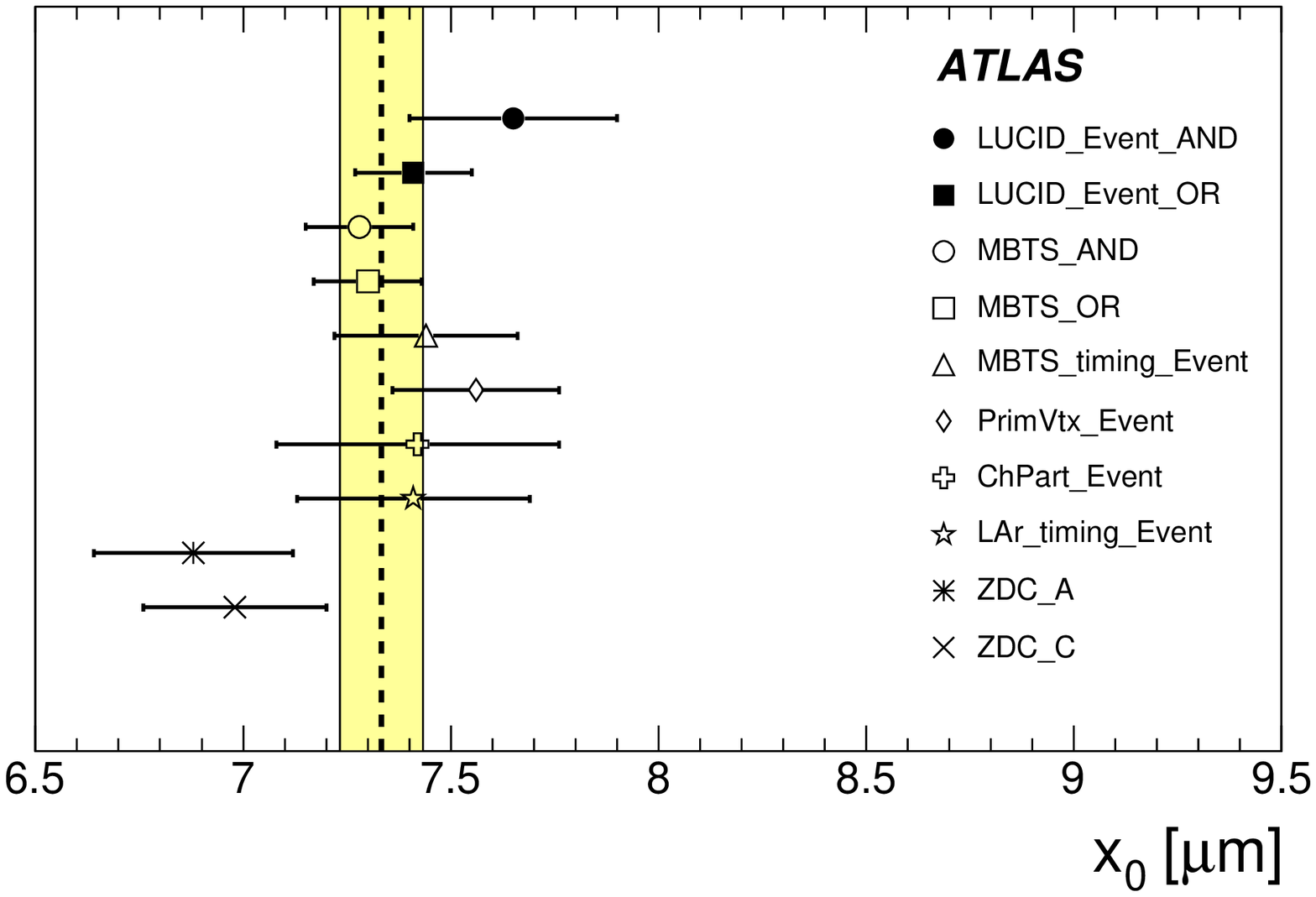,width=0.48\textwidth}
}
\hfil
\subfigure[]{
\epsfig{file=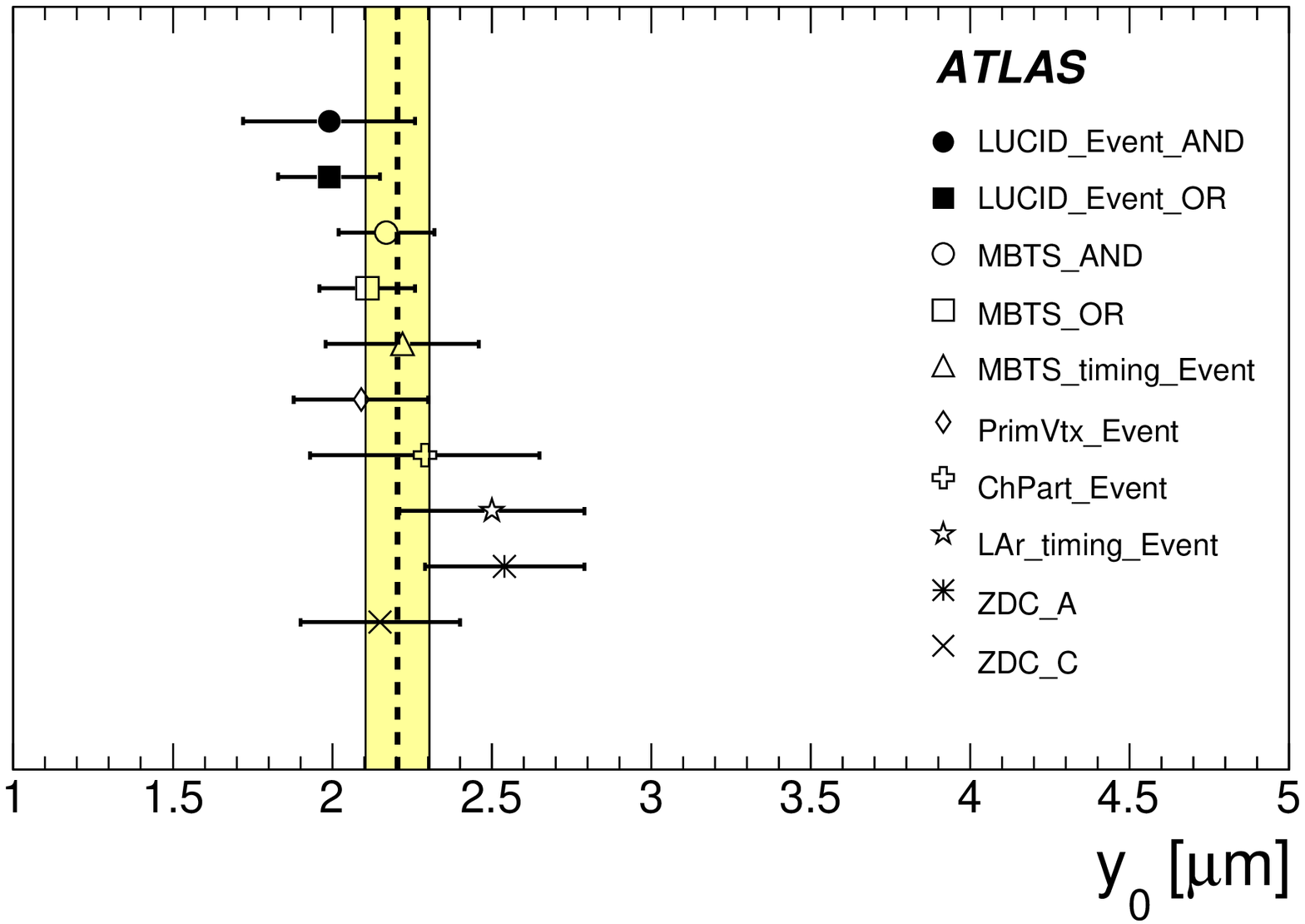,width=0.48\textwidth}
}
\caption{
Fit results for the values of 
(a) $\Sigma_x$, (b) $\Sigma_y$, (c) $x_0$ and
(d) $y_0$
obtained using different luminosity algorithms during Scan~II.
The dashed vertical line
shows the unweighted average of all the algorithms.
The
shaded bands indicate $\pm 0.5\%$ deviations from the mean 
for (a) and (b) and
$\pm 0.1\mu $m deviations from the mean for (c) and (d).
In all cases, 
the uncertainties on the points are the statistical errors reported by
the {\it vdM} fit.  Uncertainties for different algorithms
using the same detector are correlated.
\label{fig:vdMCapSigmaAllMethods}
}
\end{figure*}

Summaries of the relevant fit parameters for the three scans
are presented in Tables~\ref{tab:VdMFits-I} through~\ref{tab:VdMFits-III}
in Appendix~A.
Because the emittance during Scan~I was different from that
during Scans~II and~III, the values of $\Sigma_x$ and $\Sigma_y$ are not
expected to be the same for the first and the later scans. Furthermore, because the 
beam currents were lower in Scan~I, the peak luminosities for this scan are
lower than for the later scans. These tables, as well as 
Fig.~\ref{fig:vdMCapSigmaAllMethods}, show that the 
mean position and $\Sigma$ 
for a given scan are consistent within statistical uncertainties amongst all 
algorithms. These data also indicate several potential sources of systematic 
uncertainty.  First, the fitted position
of the peak luminosity deviates from zero by as much as $7\;\mu$m, indicating 
that the beams may not have been properly centred before the start of the scan.  
Second, in  scans~II and III, the peak luminosities for the horizontal and vertical 
scans, as measured with a single algorithm, show a systematic difference of 
as much as 5\%\ (with a lower rate observed in the vertical scan for all algorithms). This 
systematic dependence may indicate a level of irreproducibility in the scan setup. 
The effect of these systematic uncertainties on the luminosity calibration is 
discussed in Section~\ref{sec:VdMsystematics}.

Calibration of the absolute luminosity from the beam scans 
uses the following expression for $\sigma_{vis}$:
\begin{equation}
\sigma _{vis}  = \frac{{R^{MAX} }}{{\mathcal L}^{MAX}} = R^{MAX}
\frac{{2\pi \Sigma _x \Sigma _y }}{{n_b f_r (n_1 n_2)_{MAX} }}\label{svis}
\end{equation}
where $R^{MAX}$ and ${\mathcal L}^{MAX}$ are, respectively, the value of $R_{sp}$ 
and the absolute luminosity (inferred from the measured machine parameters) when 
the beams collide exactly head-on. Since there are two independent measurements, 
one each for the $x$ and $y$ directions,  and each has the same statistical 
significance, the average of the two measurements is considered as the best 
estimate of $R^{MAX}$:
\begin{equation}
R^{MAX}  = \frac{1}{2}(R_x^{MAX}  + R_y^{MAX} )\label{rmax}
\end{equation}
The values of $\sigma_{vis}$ for each method and each scan are reported
in Table~\ref{tab:sigmavisperscan} in Appendix~A.
While the results of the second and third luminosity scans are compatible
within statistical uncertainties, those of the first luminosity scan
are lower by 2.7\%\ to 4.8\%  for all online algorithms, but are consistent
for the offline track-based algorithms. 
These differences again indicate possible systematic variations
occurring between machine fills and are most likely to be caused by
variations in the beam current calibration 
(see Section~\ref{sec:VdMsystematics}).

Figure~\ref{fig:vdMSpecifLumi} 
(and Table~\ref{tab:sigmavisperscan} in Appendix~A)
also report the specific
luminosity normalized to units of $10^{11}$ protons per bunch
\begin{equation}
\mathcal{L}_{spec} = 10^{22} ({\rm p/bunch})^2 
\frac{f_r}{2\pi \Sigma_x \Sigma_y}  
\end{equation}
Because the emittance of Scan~I was
different from that of Scans~II and~III, the specific luminosity of that
scan is not expected to be the same as for the later scans.
The agreement between algorithms within one scan is excellent.
This agreement demonstrates that the variation in the measured
value of $\sigma_{vis}$ with scan number for a given algorithm
is due to variations in the fitted value of $R^{MAX}$ rather than 
in the values obtained for $\Sigma$.
\begin{figure*}
\subfigure[]{
\epsfig{file=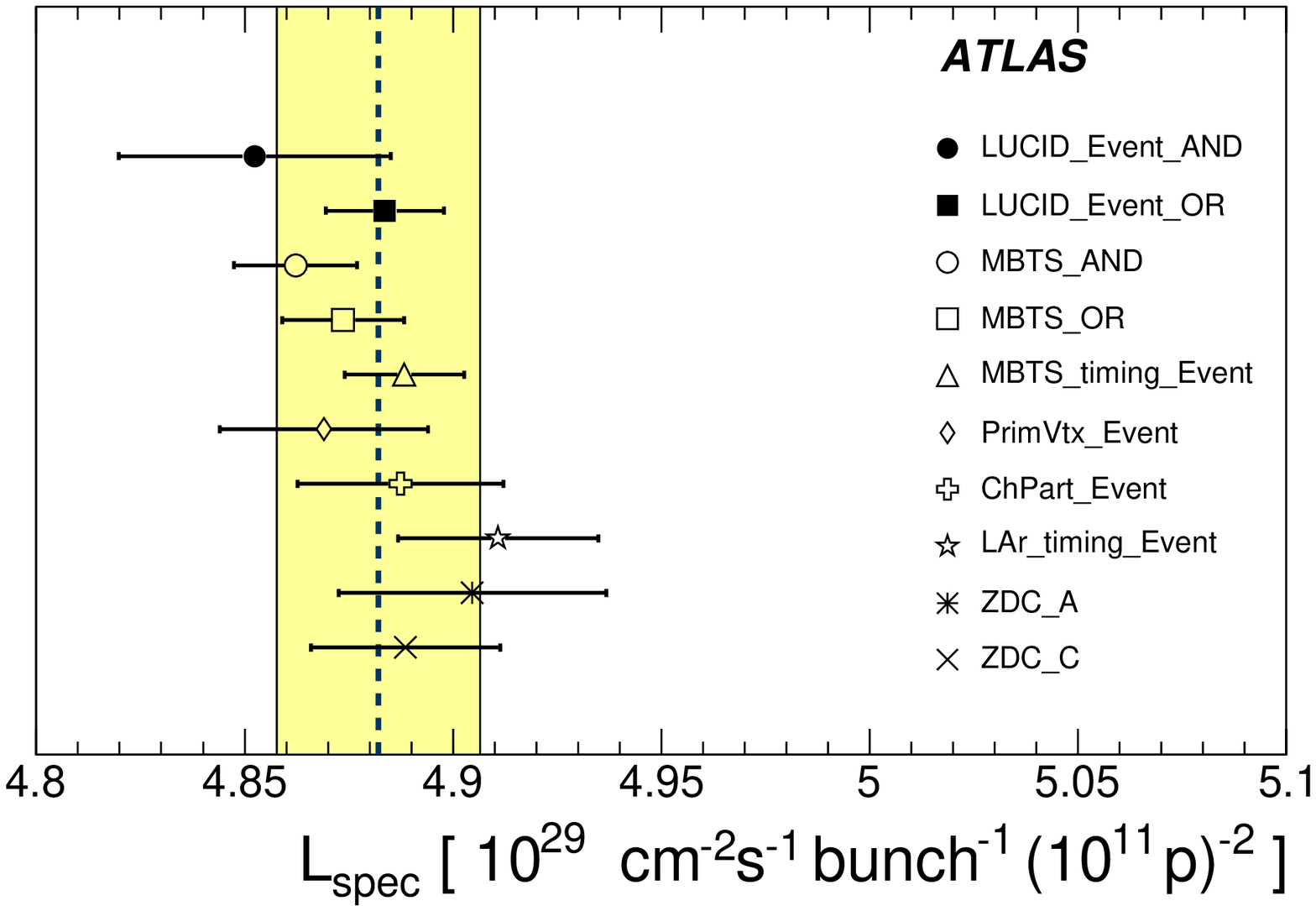,width=0.48\textwidth}
}
\subfigure[]{
\epsfig{file=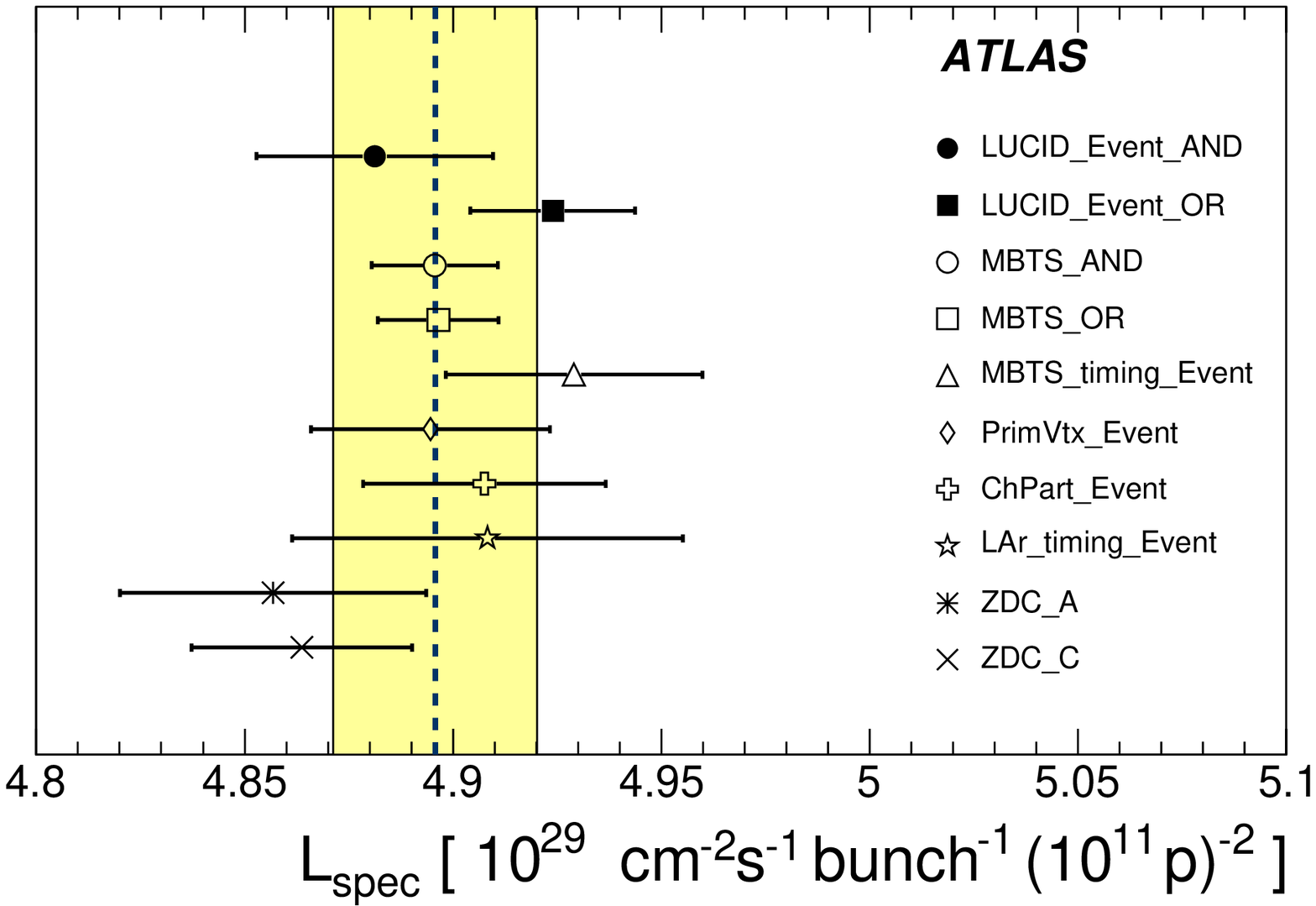,width=0.48\textwidth}
}
\caption{
Comparison of the specific luminosities obtained 
using various luminosity algorithms for (a) Scan~II and
(b) Scan~III.
The dashed lines
show the unweighted average of all algorithms; the
shaded band indicates a $\pm 0.5\% $ variation from that mean.
The uncertainties on the points are the statistical errors reported by
the \vdM\ fit.  Uncertainties for different algorithms
using the same detector are correlated.
}
\label{fig:vdMSpecifLumi}
\end{figure*}

\subsection{Systematic Uncertainties}
\label{sec:VdMsystematics}
\begin{table*}[tp]
\begin{center}
\begin{tabular}{|l|c|}
\hline
Source & Uncertainty on $\sigma_{vis}$ (\% ) \\
\hline
Beam Intensities & 10 \\
Length-Scale Calibration & 2\\
Imperfect Beam Centring & 2 \\
Transverse Emittance Growth \& Other Sources of Non-Reproducibility & 3 \\
$\mu $ Dependence & 2\\
Fit Model & 1 \\
\hline
Total & 11 \\
\hline
\end{tabular}
\end{center}
\caption{Summary of systematic uncertainties on the visible cross sections
obtained from beam scans.  Because $\sigma_{vis}$ is used to determine
the absolute luminosity (see Equation~\ref{eq:calib}), these results
are also the systematic uncertainty on the beam-scan based luminosity
calibrations.
}
\label{tab:VdmSysSum}
\end{table*}
Systematic uncertainties affecting the luminosity and visible cross section 
measurements
arise from the following effects.
\begin{enumerate}
\item {\bf Beam intensities}\\
A systematic error in the measurement of the absolute bunch charge translates 
directly into an uncertainty on the luminosity calibration. 
The accuracy of the  bunch intensity measurement depends on that of 
the DCCT calibration. While  laboratory measurements indicate an rms 
absolute scale uncertainty of better than 1.2\%, 
the DCCT suffers from slow baseline drifts that are beam-, time- and 
temperature-depend\-ent. These baseline offsets can only be determined 
with no beam in the LHC. 

For the fills under consideration, the DCCT baseline was measured before 
injection, and then again after dumping the beam. The DCCT-baseline 
determination is subject to magnetic and electronic drifts that translate 
into an rms uncertainty on the total circulating charge of 
$\sim 1.15 \times 10^{9}$ protons.
Conservatively combining the uncertainty on the absolute scale and on the 
baseline subtraction linearly yields a fractional uncertainty on the 
total charge $n_{1(2)}$ in beam 1 (2) of
\begin{equation}
\frac{\sigma(n_{1(2)})}{n_{1(2)}} = \frac{1.15\times 10^{9}}{n_b n_{1(2)}}+0.012
\end{equation}
Treating the current-scale uncertainty as fully correlated between the two
beams results in a total systematic error of  $\pm 10\% $ on 
the product of bunch
currents for the running conditions summarized in Table 2. Because the
baseline correction dominates the overall bunch-charge uncertainty, and
because it drifts on the time scale of a few hours, these uncertainties are
largely uncorrelated between the first (scan I) and the second (scans
II+III) luminosity-calibration sessions.
\item {\bf Length-Scale Calibration}\\
Fits to the beam size depend on knowledge of the relative displacement between
the beams at each scan step.  Thus, any miscalibration of the beam
separation length-scale will result in a mismeasurement of the luminosity.
The desired nominal beam  separation during beam scans 
determines the magnet settings of the closed orbit bumps that 
generate the beam separation.  
The only accelerator instrumentation available for calibrating
the length-scale of the beam separation is the beam position monitor
system.  Unfortunately, 
the short-term stability and reliability of this system 
are not adequate to perform such a calibration.  
In contrast, the vertex resolution of the ATLAS Inner Detector provides
a stable and precise method of calibration.  These calibrations
were done in dedicated scans where both beams were moved in the
same direction first by $+100\ \mu$m and then by $-100\ \mu$m
from the nominal beam position, first in the horizontal and then in the
vertical direction.  The luminous beam centroid was determined using
reconstructed primary vertices.  In addition, the primary vertex event
rate was monitored to ensure that the two beams remained centred with 
respect to each other.  
The calibration constants derived for the length-scale were $(1.001\pm 0.003)$
and $(1.001\pm 0.004)$ in the horizontal and vertical directions respectively,
indicating that the scale associated with the magnet settings and that obtained
from the ATLAS Inner Detector agree to better than $0.5 \%$. The dominant 
source of uncertainty is the precision with which the two beams could be kept 
transversely aligned during the length-scale calibration scans. In addition, 
these scans consisted of only three points and extended to only 
$\pm 100\ \mu$m; therefore these data do not allow for studies of 
non-linearities, nor for checks of the calibration at the larger beam 
displacements used during the luminosity-calibration scans. Finally, if 
the transverse widths of the two beams happened to be significantly different,
the measured displacements of the luminous centroid at each scan point would
not exactly reflect the average displacement of the two beams. The combination
of these effects results in an estimated systematic uncertainty of 2\%\ on 
the length-scale calibration, in spite of the high precision of the 
calibration-scan data. 
\item {\bf Imperfect Beam Centring}\\
If the beams are slightly offset with respect to each other 
in the scan direction, there is no impact on the  
results of the luminosity scan.
However, a deviation 
from zero separation in the transverse direction orthogonal  
to that of the scan reduces the rate observed for
all the data points of that scan. The  
systematic uncertainty associated with imperfect beam centring
has been estimated by considering the maximum deviation 
of the peak position (measured in terms of the nominal beam separation)
from 
the nominal null separation that was 
calibrated through the re-alignment of the 
beams at the beginning of that scan. This 
deviation is translated into an expected decrease in rate and therefore in 
a systematic uncertainty 
affecting the measurement of the visible cross section. 
A systematic uncertainty of 2\%\ is assigned.

\item {\bf Transverse Emittance Growth and Other Sources of Non-reproducibility}\\
Wire-scanner measurements of the transverse emittances of the LHC beams were
performed at regular intervals during the luminosity-scan sessions, yielding
measured emittance degradations of roughly 5-10\% per beam and per plane between
the beginning and the end of the luminosity-calibration
sessions~\cite{bib:emittanceSimon}. This emittance growth causes a progressive
increase of the transverse beam sizes (and therefore of $\Sigma_x$ and $\Sigma_y$),
leading to a $\sim 2\%$ degradation of the specific luminosity between the first and the 
last scan within one session. This luminosity degradation, in turn, should be reflected in a
variation over time of the specific rates $R_x^{MAX}$ and $R_y^{MAX}$
(Eq.~\ref{rmax}). 
A first potential bias arises because if the time dependence of $\Sigma_x$ and
$\Sigma_y$ during a scan is not taken into account, the emittance growth may
effectively distort the luminosity-scan curve.
Next, and because the horizontal and vertical scans were separated in time,
uncorrected emittance growth may induce inconsistencies in computing the luminosity
from accelerator parameters using Eq.~\ref{svis}.
The emittance growth was estimated independently from the wire-scanner data, and
by a technique that relies on the relationship, for Gaussian beams,
between $\Sigma$, the single-beam sizes $\sigma_1$ and $\sigma_2$
and the transverse luminous size $\sigma_L$ (which is measured using the spatial
distribution of  primary vertices)~\cite{bib:CONF027}:
\begin{eqnarray}\nonumber
\Sigma & = & \sqrt {\sigma_{1}^2+\sigma_{2}^2} \\
\frac{1} {\sigma_{L}} & = &
\sqrt{
\frac{1}{\sigma_1^2} +
\frac{1}{\sigma_2^2} } 
\end{eqnarray}
Here the emittance growth is taken from the measured
evolution of the transverse luminous size during the fill.
The variations in both $\Sigma$ and $R^{MAX}$ (which should in principle
cancel each other when calculating the visible cross-section) were then predicted 
from 
the two emittance-growth estimates, and compared to the luminosity-scan results. 
While the predicted variation of $\Sigma$ between consecutive scans is very small 
(0.3~-~0.8\,$\mu$m) and well reproduced
by the data, the time evolution of $R^{MAX}$ displays irregular deviations from the
wire-scanner prediction of up to 3\%, suggesting that at least one additional
source of non-reproducibility is present.
Altogether, these estimates suggest that a $\pm 3$\% systematic uncertainty on
the luminosity calibration be assigned to emittance growth and unidentified causes
of non-reproducibility.

\item {\bf $\mu $-Dependence of the Counting Rate} \\
All measurements have been corrected for 
$\mu$ dependent non-linearities.
Systematic uncertainties on the predicted
counting rate as a function of $\mu$ have been studied using Monte Carlo
simulations, where the efficiency (or equivalently  $\sigma_{vis}$)
have been varied.  For $\mu<2$ the uncertainty is estimated to be $< 2\%$, 
as illustrated in Fig.~\ref{fig:MCvsParameterizationFnMu}.

\item {\bf Choice of Fit Model} \\
For all methods, fits of the scan data to the default function 
(double Gaussian with common mean plus constant background for the
online algorithms and double Gaussian for the background-free offline
algorithms) have $\chi^2$ per degree of freedom
values close to 1.0, indicating that the fits
are good.  The systematic uncertainty due to this choice of fit function
has been estimated by refitting the offline data using a cubic spline
as an alternative model.  The value of $\sigma_{vis}$ changes by 
approximately $1\% $.
\end{enumerate}
A summary of the systematic uncertainties is presented in 
Table~\ref{tab:VdmSysSum}.  The overall uncertainty of 11\%\ is
dominated by the measurement of the beam intensities. 
At least some portion of this uncertainty is common to interactions points
1 (ATLAS) and 5 (CMS); the size of this correlated uncertainty
remains under study.

\section{Internal Consistency of Luminosity Measurements}
\label{sec:Cnstcy}
\begin{figure}
\begin{center}
\subfigure[]{
\epsfig{file=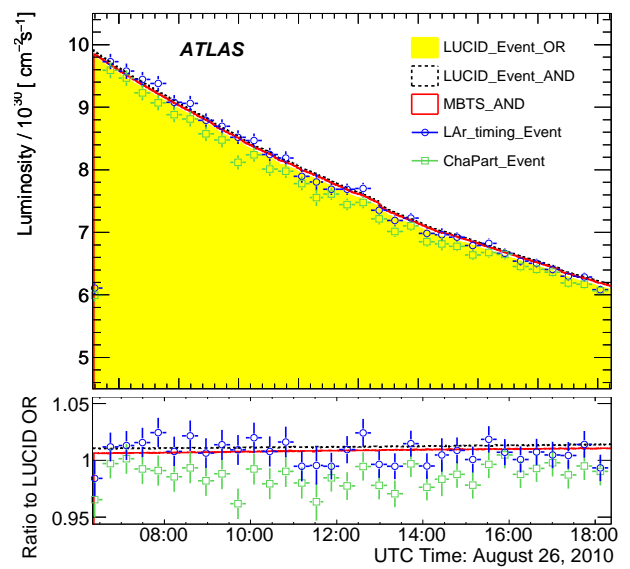,width=0.45\textwidth}
}
\subfigure[]{
\centerline{\epsfig{file=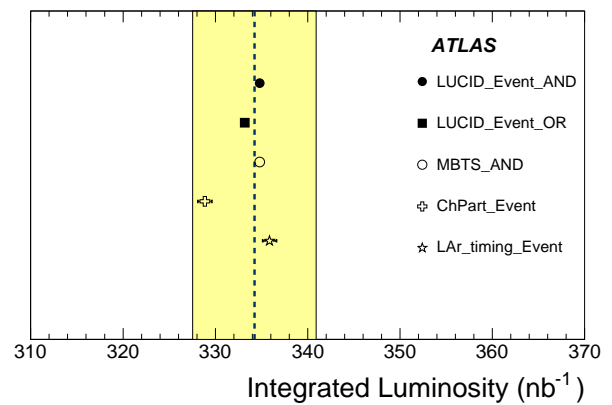,width=0.45\textwidth}}
}
\end{center}
\caption{(a) 
ATLAS instantaneous luminosity for Run 162882, as measured using several 
algorithms.  Each curve is independently normalized using the {\it vdM} calibration 
obtained for that algorithm.  The inset at the bottom shows the ratio of the
luminosity obtained with each algorithm to that obtained with \lucidEventOr.
The statistical uncertainties for the online algorithms (\lucidEventOr, 
\lucidEventAnd\ and \mbtsoneone) are negligible.  Statistical uncertainties
for the offline algorithms (\LARtiming\ and \chpart) are displayed.
(b) Comparison of the integrated luminosity obtained for Run 162882 for each
of the algorithms shown above, together with the statistical 
uncertainties on the measurements.  The dotted line shows the
weighted mean of all the algorithms.  The shaded band indicates a $\pm 2\%$
deviation from that mean.}
\label{fig:compareLumiVDMNormalization}
\end{figure}

It is possible to test the consistency of the {\it vdM} calibrations by
comparing the luminosities obtained using different luminosity detectors and/or 
algorithms. Figure~\ref{fig:compareLumiVDMNormalization} shows the instantaneous 
luminosities obtained by various algorithms for Run 162882\footnote{The
bunch-by-bunch luminosity for \lucidEventOr\ averaged over the full run is shown in 
Fig.~\ref{fig:LUCIDLumiPerBCID}.}, each normalized using the calibration 
extracted from its {\it vdM} scan data. The absolute luminosities agree to better 
than $2\% $; the relative luminosities track each other over time to within the 
statistical fluctuations. 
Over most of the 2010 \pp\ run, \lucidEventOr\  was chosen as the preferred
offline algorithm because its pile-up correction was well-understood, its
statistical power was adequate and backgrounds for this algorithm were low. 

\begin{figure}
\begin{center}
\epsfig{file=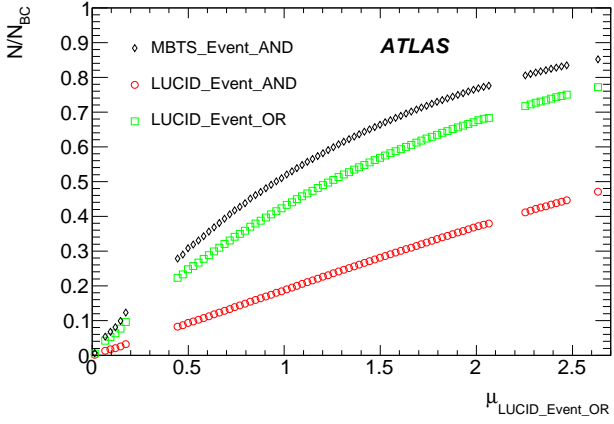,width=0.48\textwidth}
\end{center}
\caption{Fraction of bunch crossings containing a detected event
(N/N$_{BC}$) for several
algorithms, as a function of $\mu_{\tinylucidEventOr}$.}
\label{fig:ratesVsMuOR}
\end{figure}

\begin{figure}
\begin{center}
\epsfig{file=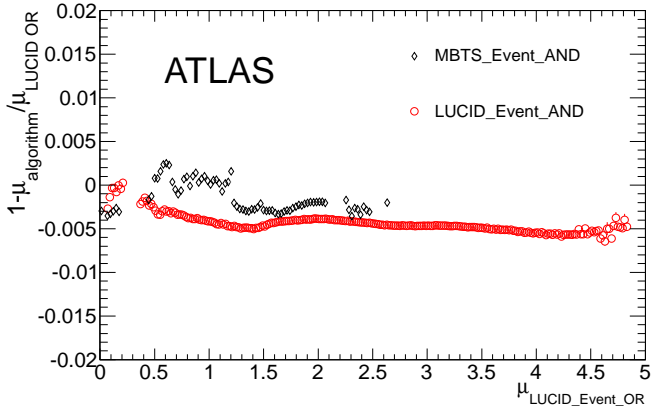,width=0.48\textwidth}
\end{center}
\caption{Fractional deviation  
of the average value of $\mu$ obtained with the
\mbtsoneone\ and \lucidEventAnd\ algorithms 
with respect to the \lucidEventOr\ algorithm as a function of 
$\mu$ obtained with \lucidEventOr.}
\label{fig:lumiRatioVsMuOR}
\end{figure}

Comparing the residual $\mu$-dependence (if any) of the measured luminosity across 
multiple detectors and algorithms probes the consistency of the pile-up correction 
procedures described in Section~\ref{subsec:mudependence}. 
Figure~\ref{fig:ratesVsMuOR} shows, for some of the LUCID and MBTS algorithms, the 
raw counting rate as a function of 
the average number of 
inelastic interactions per BC measured by \lucidEventOr\  using the prescription of 
Section~\ref{subsubsec:muEvOr}. 
Non-linearities
are apparent (as expected) for the \lucidEventAnd, \lucidEventOr\ 
and \mbtsoneone\ algorithms. If the para\-metrizations of 
Section~\ref{subsec:mudependence} are correct, however, then the ratio of the 
luminosities determined using the different algorithms should be independent of 
$\mu$.  Figure~\ref{fig:lumiRatioVsMuOR} shows that the values
of $\mu$ obtained with the \lucidEventAnd\
and \mbtsoneone\ algorithms 
remain within $\pm 1\%$ of that measured using the 
\lucidEventOr\ algorithm over the range $0<\mu<2.5$.
Comparisons of the \lucidEventOr\ and \lucidEventAnd\ algorithms
demonstrate agreement up to $\mu=5$, the highest value of $\mu$ obtained
during the 2010 LHC run.
No results are presented beyond $\mu = 2.5$ for the MBTS because during the
corresponding data-taking period the short spacing between consecutive LHC
bunches made the MBTS luminosity measurement unreliable. Possible causes
include the long duration of the analog pulse, 
saturation effects following large energy deposits,
time jitter introduced by the
electronics used at the time, and afterglow background. 

\section{Comparison with Monte Carlo Generators}
\label{sec:MC}

Because the \vdM\ method does not require knowledge of the 
inelastic cross section nor of the detector acceptance, 
the values of \sigmavis\ obtained from the beam scans
can be used to test the accuracy of the predictions of Monte Carlo event 
generators.  Such predictions suffer from several theoretical uncertainties.
First, because the
\pp\ inelastic cross section has not been measured at 7 TeV,
the generators obtain \sigmainel\ by extrapolating
from lower energy.  Results of this extrapolation depend on the
functional form  used.  
The \pythia\ and \phojet\ generators,
for example, predict
values for \sigmainel\ that differ by 6.6\%.  Second,
the generators must separately model the non-diffractive (ND), 
single-diffractive (SD) and double-diffractive (DD) components
of the cross section.  There exists no unique prescription for
classifying events as diffractive or non-diffractive and
no calculation of the cross sections from first principles.
Typical uncertainties associated with such classifications
are illustrated in Table~\ref{tab:mcCrossSection}.
The fraction of \sigmainel\ corresponding to ND events is 68\%\ in \pythia\
and 81\%\ in \phojet, while the DD fractions
are 13\%\ and 5\%\ respectively.   Finally, there are significant 
uncertainties on the modeling of the predicted
multiplicity-, $\pT$- and $\eta$- distributions for particles produced 
in soft \pp\ interactions, particularly for the poorly 
constrained diffractive components.  Differences in these distributions
will affect the efficiencies for events to pass the selection criteria
of a specific luminosity algorithm.

Within the framework of Monte Carlo generators, \sigmavis\ is calculated using the 
expression
\begin{equation}
\sigmavis = 
\epsilon_{ND}\sigma_{ND}+\\
\epsilon_{SD}\sigma_{SD}+\epsilon_{DD}\sigma_{DD}
%+\epsilon_{CD}\sigma_{CD}}
\label{eq:sigmavis}
\end{equation}
where $\epsilon_{process}$ are the efficiencies and $\sigma_{process}$ 
the cross sections for the individual inelastic processes (ND, SD and DD).
Table~\ref{tab:accept7TeV}  shows the predicted efficiencies for observing ND, SD 
and DD events using either \pythia\ (with the default ATLAS MC09 
tune\cite{bib:pythiamc09}) or \phojet, for some of the algorithms described in 
Section~\ref{sec:methods}. In general, the \phojet\ predictions are about 
15\%\ to 20\%\ 
higher than those obtained with \pythia. One exception is \lucidEventAnd\ which is 
less sensitive to the diffractive processes: here the two generators agree to 
within 5\% overall. Additional systematic uncertainties on these predictions, 
associated with the modeling of the detector response in the simulation, are 
algorithm- and trigger-dependent and vary from $2.2\%$~for \mbtsone\ to 
$6\%$~for\\
\lucidEventAnd.
%typically amount to $\sim \pm 4\%$.  
\par
\begin{figure*}
\subfigure[]{
\epsfig{file=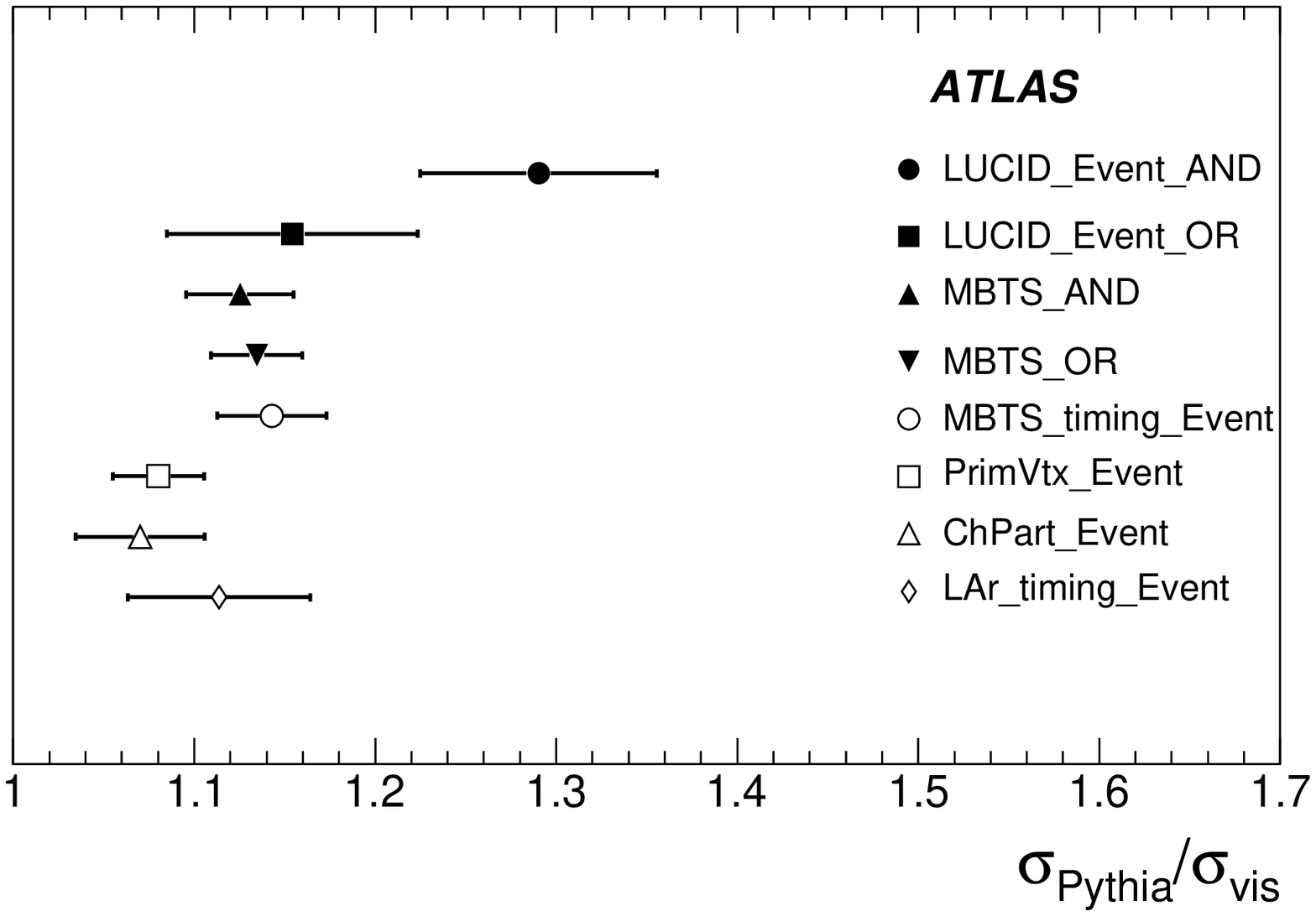,width=0.48\textwidth}
}
\subfigure[]{
\epsfig{file=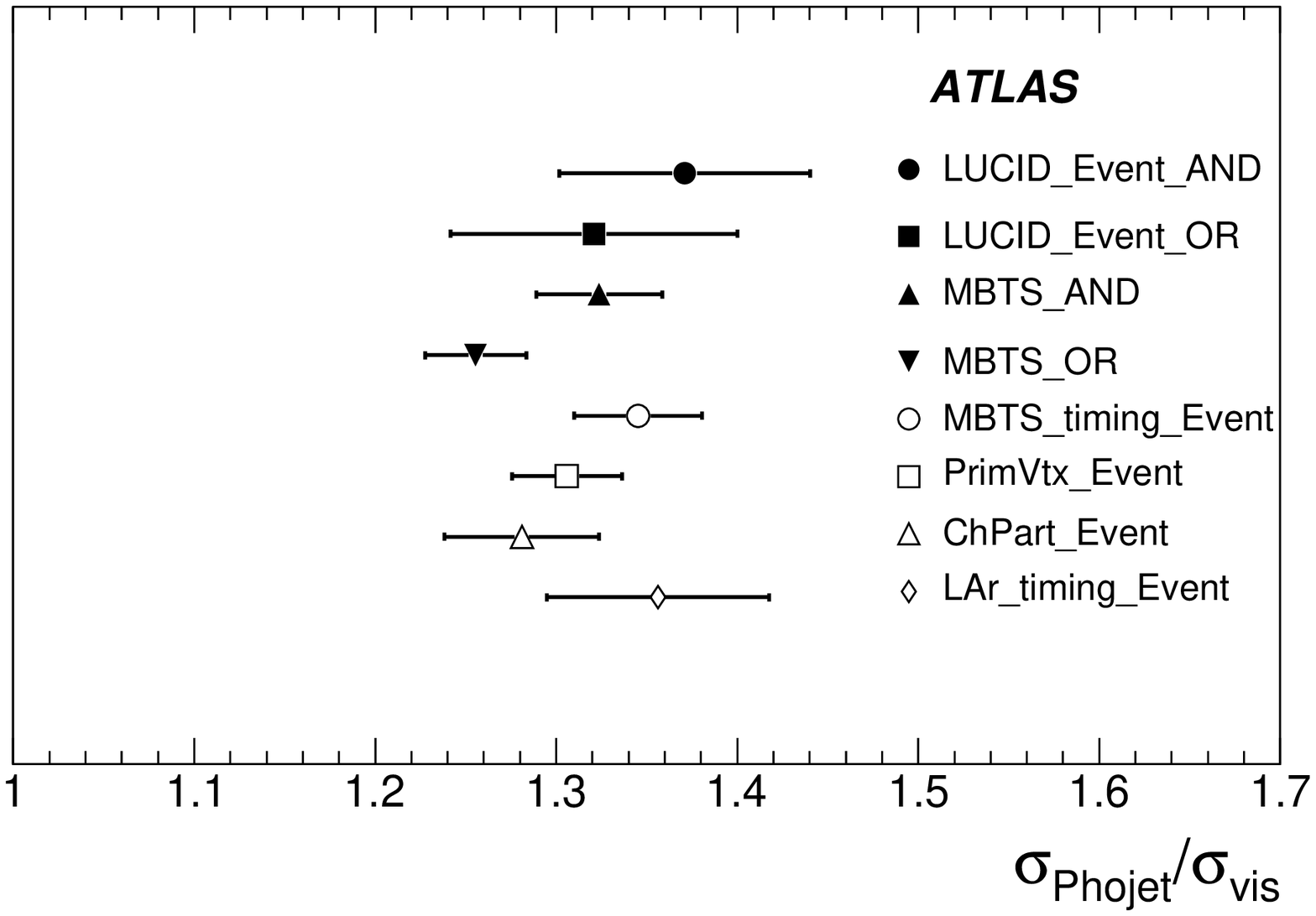,width=0.48\textwidth}
}
\caption{Comparison of the measured values of \sigmavis\ for several
algorithms to the (a) \pythia\ and (b) \phojet\ predictions.
%%The dashed vertical line shows the unweighted average over all the
%%algorithms shown; the shaded band indicates the rms deviation.
The errors on the points are the systematic
uncertainties due to possible inaccuracies in
modeling the detector response.
The uncertainties for different algorithms
using the same detector are correlated.
The 11\%\ uncertainty on the \vdM\ calibration of the luminosity, 
which is 100\%\ correlated among algorithms, 
is {\it not} included in the error bars.
}
\label{fig:compareToMC}
\end{figure*}

As noted in Section~\ref{sec:VdmResults}, there is a systematic difference between 
the values of $\sigma_{vis}$ obtained from the first scan and those based on the 
second and third scans. In reporting our best estimate of the measured visible 
cross sections, we chose to average the results of the first scan with the average of 
the second and third scans. Comparisons of the \vdM\ scan measurements with the 
Monte Carlo predictions are presented in Table~\ref{tab:compareToMC} and  
Fig.~\ref{fig:compareToMC}. For a given event generator, the comparisons exhibit 
an RMS spread of 4 to 5\%; on the average, the \pythia\ (\phojet) predictions are  
15\%\ (33\%) higher than the data. Given the 11\%\ systematic uncertainty on the 
\vdM\ calibration, which is correlated across all algorithms, \pythia\ agrees with 
the data at the level of 1.5$\sigma$, while \phojet\ and the data deviate at the 
3$\sigma$ level.

\begin{table}[tbp]
\begin{center}
\begin{tabular}{|l|r|r|}
\hline
\multicolumn{3}{|c|}{Cross Section at $\sqrt s = 7$ TeV}\\
\hline
Process & \pythia\ & \phojet \\
        & \multicolumn{1}{|c|}{(mb)} & \multicolumn{1}{|c|}{(mb)} \\
\hline
non-diffractive (ND) & 48.5  &  61.6 \\
single-diffractive (SD) & 13.7 & 10.7 \\
double-diffractive (DD) & 9.3 & 3.9 \\
%central-diffractive & n.a. & 1 \\
\hline
Total: & 71.5 & 76.2  \\
\hline
\end{tabular}
\end{center}
\caption{Predicted inelastic \pp\ cross sections 
at $\sqrt s = 7$~TeV for \pythia\ and for \phojet.
A small ($\sim 1$ mb) contribution from double-pomeron
processes (``central diffraction'') was not included in the \phojet\
cross section.}
\label{tab:mcCrossSection}
\end{table}
%%%%% 
\begin{table*}[tbp]
\begin{center}
\begin{tabular}{|l|r|r||r|r|}
\hline
 & \multicolumn{2}{|c||}{\lucidEventOr } & \multicolumn{2}{|c|}{\lucidEventAnd }\\
\hline
\hline
Process & \multicolumn{2}{|c||}{Efficiency (\%\ ) } & 
\multicolumn{2}{|c|}{Efficiency (\%\ )}\\
\hline
       & \pythia\ MC09 & \phojet\ & \pythia\ MC09 & \phojet\ \\
\hline
ND & 79.7 & 73.7 & 30.8 & 24.9 \\
SD & 28.7 & 44.3 & 1.3 & 2.4 \\
DD & 39.9 & 62.0 & 4.3 & 14.6 \\
\hline
\hline
$\sigma_{vis}$ (mb) & 46.4 & 53.1 & 16.0 & 17.0\\
\hline
\multicolumn{5}{c}{ } \\
\hline
 & \multicolumn{2}{|c||}{\mbtsOFF\ } & \multicolumn{2}{|c|}{\LARtiming } \\
\hline
\hline
Process &  \multicolumn{2}{|c||}{Efficiency (\%\ )} &  
\multicolumn{2}{|c|}{Efficiency (\%\ )} \\
\hline
        & \pythia\ MC09 & \phojet\ &  \pythia\ MC09 & \phojet\ \\
\hline
ND & 97.4 & 97.9 & 96.0 & 94.3 \\
SD & 41.3 & 44.3 & 21.4 & 27.9  \\
DD & 50.8 & 68.1 & 25.9 & 53.6\\
\hline
\hline
$\sigma_{vis}$ (mb) & 57.6 & 67.8 & 51.9 & 63.2 \\
\hline
\multicolumn{5}{c}{ } \\
\hline &
 \multicolumn{2}{|c||}{\chpart} &
 \multicolumn{2}{|c|}{\mbtspvtx}\\
\hline
\hline
Process & \multicolumn{2}{|c||}{Efficiency (\%\ )} & 
\multicolumn{2}{|c|}{Efficiency (\%\ )} \\
\hline
        & \pythia\ MC09 & \phojet\ & \pythia\ MC09 & \phojet\ \\
\hline
ND & 85 & 80 & 97.8 & 99.2 \\
SD & 36 & 36 & 43.9 & 56.9\\
DD & 36 & 41 & 47.8 & 70.7 \\
\hline
\hline
$\sigma_{vis}$ (mb) & 45.7 & 54.7 & 57.9 & 70.0 \\
\hline
\end{tabular}
\end{center}
\caption{Efficiencies at $\sqrt s =7$~TeV for several of the  
luminosity methods described in Section~\ref{sec:methods}.
The predicted visible cross sections $\sigma_{vis}$ are obtained using 
Equation~\ref{eq:sigmavis}, the efficiencies in the present table and the 
cross sections in Table~\ref{tab:mcCrossSection}.
}
\label{tab:accept7TeV} 
\end{table*}
%%%%%%%
\begin{table*}
\begin{center}
{%\small
\begin{tabular}{|l|c||c|c||c|c|}
\hline
 & & & & &  \\
Algorithm & $\sigma_{vis}$ & $\sigma_{vis}^{\pythia }$ &
$\frac{\sigma_{vis}^{\pythia }}{\sigma_{vis}}$ &
$\sigma_{vis}^{\phojet}$ &
$\frac{\sigma_{vis}^{\phojet }}{\sigma_{vis}}$ \\
& (mb) & (mb) &  & (mb) &  \\
\hline
\lucidEventAnd\ & $12.4 \pm 0.1$ & $16.0 \pm 0.8$ & $1.29 \pm 0.07$ & $17.0\pm 0.9 
$ & $1.37\pm 0.07$ \\
\lucidEventOr\ & $40.2 \pm 0.1 $ & $46.4 \pm 2.8$ & $1.15 \pm 0.07$ & $53.1 \pm 
3.2$ & $1.32\pm 0.08$ \\
\mbtsoneone\ & $51.9 \pm 0.2 $ & $58.4 \pm 1.5$ & $1.13 \pm 0.03$ & $ 68.7 \pm 1.8$ 
& $1.32\pm 0.03$ \\
\mbtsone\  &  $58.7 \pm 0.2 $ & $ 66.6 \pm 1.5$ & $1.13 \pm 0.03$ & $73.7 \pm 1.6$ 
& $1.26\pm 0.03$\\
\mbtsOFF\  & $50.4 \pm 0.2 $ & $57.6 \pm 1.3$ & $1.14 \pm 0.03$ & $67.8\pm 1.8 $ & 
$1.35\pm 0.04$ \\
\mbtspvtx\ & $53.6 \pm 0.2 $  & $57.9 \pm 1.3$ & $1.08\pm 0.03$  & $70.0\pm 1.6 $ &  
$1.31\pm 0.03$ \\
\chpart \ & $42.7 \pm 0.2$ &  $45.7 \pm 1.7$ & $1.07 \pm 0.04$ & $54.7\pm 2.0$ & 
$1.28\pm 0.05$\\
\LARtiming \ & $46.6 \pm 0.2$ & $51.9\pm 2.3$  & $1.11\pm 0.05$ & $63.2\pm 2.9$  & 
$1.36\pm 0.06$ \\
\hline
\end{tabular}
}
\end{center}
\caption{Comparison of the visible cross sections determined from beam scans 
($\sigma_{vis}$) to the predictions of the \pythia\ and \phojet\ Monte Carlo
generators. The ratio of prediction to measurement is also shown. The errors 
affecting the measured visible cross sections are statistical only. The errors on 
the \pythia\ and \phojet\ visible cross sections are obtained from
the systematic 
uncertainty associated with modeling the detector response. 
These uncertainties are 
fully correlated, row by row, between \pythia\ and \phojet; they are fully 
correlated between the two LUCID algorithms, 
and highly correlated for the five 
MBTS-triggered algorithms (\smmbtsoneone, \smmbtsone, \smmbtsOFF, \smmbtspvtx\ and 
\smchpart).
The fully correlated 11\%\ systematic uncertainty on visible cross sections, that 
arises from the \vdM\ calibration, is not included in the errors listed in this 
table. }
\label{tab:compareToMC}
\end{table*}

\section{Conclusions}
\label{sec:conclusions}
Measurements of the LHC luminosity have been performed by ATLAS 
in proton-proton
collisions  at 
$\sqrt{s} = 7$ TeV using multiple detectors and algorithms.  
The absolute luminosity calibrations obtained using beam-separation
scans suffer from a $\pm 11\%$ systematic uncertainty, that is 
dominated by the uncertainty in the bunch intensities
and is therefore highly correlated across all methods. 
For a given bunch luminosity, i.e. for a fixed value of $\mu$ 
(the average number of inelastic \pp\ interactions per crossing), 
the absolute luminosities obtained using different
detectors and algorithms agree to within $\pm 2\% $. In addition, 
the luminosities from these methods track each other within better than $2\%$ over 
the range $0 < \mu < 2.5$.
The visible cross sections obtained from the beam scan calibrations also have
a systematic uncertainty of 11\%\ and are lower than those predicted by 
PYTHIA (PHOJET) by about 15 \%\ (33\%).

\section{Acknowledgements}

We wish to thank CERN for the efficient commissioning and operation of the LHC during this initial high-energy data-taking period as well as the support staff from our institutions without whom ATLAS could not be operated efficiently.
We would like, in addition, to extend special thanks to our LHC colleagues
H. Burkhardt, M. Ferro-Luzzi, S. M. White, as well as to the LHC
beam-instrumentation team, for their crucial contributions to the
absolute-luminosity calibration reported in this paper.

We acknowledge the support of ANPCyT, Argentina; YerPhI, Armenia; ARC, Australia; BMWF, Austria; ANAS, Azerbaijan; SSTC, Belarus; CNPq and FAPESP, Brazil; NSERC, NRC and CFI, Canada; CERN; CONICYT, Chile; CAS, MOST and NSFC, China; COLCIENCIAS, Colombia; MSMT CR, MPO CR and VSC CR, Czech Republic; DNRF, DNSRC and Lundbeck Foundation, Denmark; ARTEMIS, European Union; IN2P3-CNRS, CEA-DSM/IRFU, France; GNAS, Georgia; BMBF, DFG, HGF, MPG and AvH Foundation, Germany; GSRT, Greece; ISF, MINERVA, GIF, DIP and Benoziyo Center, Israel; INFN, Italy; MEXT and JSPS, Japan; CNRST, Morocco; FOM and NWO, Netherlands; RCN, Norway;  MNiSW, Poland; GRICES and FCT, Portugal;  MERYS (MECTS), Romania;  MES of Russia and ROSATOM, Russian Federation; JINR; MSTD, Serbia; MSSR, Slovakia; ARRS and MVZT, Slovenia; DST/NRF, South Africa; MICINN, Spain; SRC and Wallenberg Foundation, Sweden; SER,  SNSF and Cantons of Bern and Geneva, Switzerland;  NSC, Taiwan; TAEK, Turkey; STFC, the Royal Society and Leverhulme Trust, United Kingdom; DOE and NSF, United States of America.  

The crucial computing support from all WLCG partners is acknowledged gratefully, in particular from CERN and the ATLAS Tier-1 facilities at TRIUMF (Canada), NDGF (Denmark, Norway, Sweden), CC-IN2P3 (France), KIT/GridKA (Germany), INFN-CNAF (Italy), NL-T1 (Netherlands), PIC (Spain), ASGC (Taiwan), RAL (UK) and BNL (USA) and in the Tier-2 facilities worldwide.

\clearpage
%
%%%%%%%%%%%%%%%%%%%%%%%%%%%%%%%%%%%%%%%%%%%%%%%%%%%%%%%%%%%%%%%%%%%%%%%%%%%%%%%
% Appendix
%%%%%%%%%%%%%%%%%%%%%%%%%%%%%%%%%%%%%%%%%%%%%%%%%%%%%%%%%%%%%%%%%%%%%%%%%%%%%%
\onecolumn
\appendix
\section{Fits to Beam Scan Data}
\label{sec:appendix}
%%%\onecolumn
This appendix presents results of the fits to vdM scan data for all
scans and all algorithms.
{\small
\begin{table}[h]
\centering
\begin{tabular}{|l||c|c|c|c|c|}
\hline
Algorithm & Mean Position & $\Sigma $ &
Background & $R^{MAX}$ & $\chi^2/DOF $\\
 &  $ (\mu m)$ &  $(\mu m)$ &    (Hz) &    (Hz) &  \\
\hline
\multicolumn{6}{|c|}{Horizontal Scan~I} \\
\hline
\smlucidEventAnd & $-1.12 \pm$ 0.46 & 47.40 $\pm$ 0.56 & 0.01 $\pm$ 0.04 &
 75.6 $\pm$  1.1 &  0.9 \\
\hline
\smlucidEventOr &  $-1.58 \pm$ 0.25 & 47.27 $\pm$ 0.29 & 0.06 $\pm$0.04 &
 247.8 $\pm$     2.0 &        0.5 \\
\hline
%%\smlucidHitAnd\ & $-1.22 \pm 0.46$ & $ 47.19 \pm 0.52 $  & $0.053 \pm 0.15$  & $342.0 \pm 4.9$  & 1.4 \\
%%\hline
%%\smlucidHitOr\ & $-1.38 \pm 0.25$ & $ 47.30 \pm 0.28 $  & $0.142 \pm 0.09$  & $696.43 \pm 5.6 $  & 1.3 \\
\hline
\smmbtsoneone\ & $-1.85 \pm$ 0.25 & 47.33 $\pm $ 0.25 & 0.03 $\pm$ 0.04 &
319.0 $\pm$    2.3 & 0.8 \\
\hline
\smmbtsone\  &    $-2.05 \pm$ 0.24 &  47.30 $\pm$ 0.26 & 1.01 $\pm$ 0.11 &
361.7 $\pm$ 2.6 &  1.0 \\
\hline
\smmbtsOFF\  & $-1.66\pm 0.26$ & $47.05 \pm  0.26$ & N/A & $306.8 \pm  1.6$ & 1.0 
\\
\hline
\smmbtspvtx\ & $-1.7 \pm$ 0.2 & $ 47.26 \pm 0.25 $  & N/A & $329.7 \pm 1.6$  & 0.8 
\\
\hline
\smchpart\      & $-1.67 \pm 0.3$  & $ 47.3 \pm 0.3 $  & N/A & $253.2\pm 1.6\
$  & 0.8 \\
\hline
\smLARtiming   & $-1.44 \pm 0.27$ & $ 47.0 \pm 0.3 $  & N/A & $290.6 \pm 1.\
9$  & 0.5
\\
%\hline
%\smbcmEventAnd\ & $-4.02 \pm 2.96$  & $ 47.27 $ (fixed) & $0.018 \pm 0.011$  & $0.70 \pm 0.04$  & 0.3 \\
\hline
\smbcmEventOr\  & $-2.33 \pm 1.42$  & $ 47.27 $ (fixed)  & $7.5 \pm 0.20$  & $26.98 
\pm 0.89$  & 0.6 \\
%\hline
%\smbcmEventXOr\ & $-4.61 \pm 2.14$ & $ 47.27 $ (fixed) & $4.825 \pm 0.16$  & $13.67 \pm 0.68$  & 0.8 \\
\hline
\multicolumn{6}{|c|}{Vertical Scan~I} \\
\hline 
\smlucidEventAnd & $-5.04 \pm$ 0.50 & 55.52 $\pm$ 0.59 & 0.05 $\pm$ 0.03 &
75.8 $\pm$ 1.0 &  0.8 \\
\hline
\smlucidEventOr &  $-5.23 \pm$ 0.28 & 55.28 $\pm$ 0.33 & 0.16 $\pm$ 0.06 &
246.2 $\pm$ 1.9 & 1.1 \\
%%\hline
%%\smlucidHitAnd\ & $-5.41 \pm 0.60$  & $ 54.71 \pm 0.64 $  & $0.32 \pm 0.27$  & $342.9 \pm 4.7$  & 1.0 \\
%%\hline
%%\smlucidHitOr\ & $-5.63 \pm 0.33$  & $ 55.20 \pm 0.34 $  & $0.43 \pm 0.23$  & $694.7 \pm 5.5$  & 1.0 \\
\hline
\smmbtsoneone\ &  $-5.24 \pm$ 0.28 & 55.73 $\pm$ 0.30 & 0.10 $\pm$ 0.06 &
318.5 $\pm$ 2.3 & 1.2 \\
\hline
\smmbtsone\  & $-5.25 \pm$ 0.26 & 55.82 $\pm$ 0.28 & 1.08 $\pm$ 0.12 &
359.2 $\pm$ 2.5 & 1.2 \\
\hline
\smmbtsOFF\  & $-5.53\pm 0.30$ & $56.32 \pm 0.29$  & N/A & $297.8 \pm 1.4$  & 2.1  
\\
\hline
\smmbtspvtx\  &  $-5.17 \pm 0.26$   &56.28 $\pm 0.30$  & N/A & $323.0 \pm1.5$  & 
1.1 \\
\hline
\smchpart\      & $-5.61\pm$ 0.35 & $ 56.1 \pm 0.4 $  & N/A & $249.3\pm 1.6$\
  & 1.4\\
\hline
\smLARtiming  & $-5.11 \pm 0.31$  & $ 56.2 \pm 0.4 $  & N/A & $280.6 \pm 1.\
8$  & 2.1 \\
%\hline
%\smbcmEventAnd\ & $-3.52 \pm 8.64$  & $ 55.28 $ (fixed) & $0.056 \pm 0.038$ & $0.62 \pm 0.11$  & 0.3 \\
\hline
\smbcmEventOr\  & $-3.63 \pm 1.51$ & $ 55.28 $ (fixed) & $7.5 \pm 0.20$ & $27.3 \pm 
0.8$  & 0.7 \\
%\hline
%\smbcmEventXOr\ & $-2.22 \pm 2.47$  & $ 55.28 $ (fixed) & $4.97 \pm 0.17$ & $13.46 \pm 0.65$  & 0.4 \\
\hline
\end{tabular}
\caption{Summary of the relevant fit parameters for the Beam Scan~I.
For offline algorithms, the rates have been corrected for trigger prescales.
Because the rates in the BCM were low, the value of $\Sigma$ used for
the BCM was fixed to that obtained from the \lucidEventOr .
No results are presented for the ZDC, since the constant fraction 
discriminators used for the ZDC measurements were installed later
in the run.}
\label{tab:VdMFits-I}
\end{table}
}
{\small
\begin{table*}
\centering
\begin{tabular}{|l||c|c|c|c|c|}
\hline
Algorithm & Mean Position & $\Sigma $ &
Background & $R^{MAX}$ & $\chi^2/DOF $\\
 &  $ (\mu m)$ &  $(\mu m)$ &    (Hz) &    (Hz) &  \\
\hline
\multicolumn{6}{|c|}{Horizontal Scan~II} \\
\hline
\smlucidEventAnd & 7.65 $\pm$ 0.25 & 58.78 $\pm$ 0.16 & $-0.02 \pm$ 0.06 &
265.4 $\pm$ 3.0 & 1.8 \\
\hline
\smlucidEventOr & 7.41 $\pm$ 0.14 & 58.76 $\pm$ 0.08 & 0.07 $\pm$ 0.12 &
858.9 $\pm$ 2.5 & 2.0 \\
%%\hline
%%\smlucidHitAnd\ & $7.74 \pm 0.25$  & $ 58.12 \pm 0.18 $  & $-0.03 \pm 0.25 $  & $1250.5 \pm 6.7$  & 1.5 \\
%%\hline
%%\smlucidHitOr\ & $7.54 \pm 0.14 $  & $ 59.0 \pm 0.08 $  & $0.20 \pm 0.41$  & $2422.2 \pm 7.0$  & 2.2 \\
\hline
\smmbtsoneone\ & 7.28 $\pm$ 0.13 & 59.06 $\pm$ 0.09 & $-0.28 \pm$ 0.16 & 
1107.3 $\pm$ 3.1 & 0.9 \\
\hline
\smmbtsone\ & 7.30 $\pm$ 0.13 & 58.93 $\pm$ 0.09 & 1.04 $\pm$ 0.25 & 
1253.1 $\pm$ 3.6 &  1.2 \\
\hline 
\smmbtsOFF\  & $7.44\pm 0.22$  & $58.71\pm 0.23$  & N/A & $ 1087.0 \pm  4.1$ & 
1.3\\
\hline
\smmbtspvtx\  & $7.56\pm 0.20 $ &  $58.63 \pm 0.21 $ & N/A & $1133.0 \pm  4.0$   & 
1.1\\
\hline
\smchpart\      & $7.42 \pm 0.34$  & $ 58.5 \pm 0.2 $  & N/A & $869.1\pm 4.2\
$  & 1.1 \\
\hline
\smLARtiming\   & $7.41 \pm 0.28$ & $ 58.2 \pm 0.3 $  & N/A & $997.5 \pm 5.6\
$  & 1.6 \\
%\hline
%\smbcmEventAnd\  & $10.3 \pm 3.2$  & $ 58.76 $ (fixed) & $-0.027 \pm 0.061$ & $2.88 \pm 0.23$  & 0.7 \\
\hline
\smbcmEventOr\ & $6.54 \pm 0.59$  & $ 58.76 $ (fixed) & $0.313 \pm 0.083$ & $89.00 
\pm 0.95$  & 0.9 \\
%\hline
%\smbcmEventXOr\ & $5.64 \pm 0.90$ & $ 58.76 $ (fixed) & $0.113 \pm 0.043$ & $42.88 \pm 0.65$  & 1.2 \\
%\hline
%\smzdcEventAnd\  & $7.27 \pm 0.47$  & $ 57.67 \pm 1.65$  & $-0.01 \pm 0.08$ & $91.8 \pm 3.1$  & 1.1 \\
\hline
\smzdcA\ & $6.98 \pm 0.22$ & $ 59.05 \pm 0.12 $ & $0.09 \pm 0.14$ & $ 380.7 \pm 
1.8$  & 1.1 \\
\hline
\smzdcC\ & $6.88 \pm 0.24$ & $58.74 \pm 0.19  $ & $0.32 \pm 0.10$ & $370.57 
\pm 2.0$ & 0.8 \\
\hline
\multicolumn{6}{|c|}{Vertical Scan~II} \\
\hline
\smlucidEventAnd &  1.99 $\pm$  0.27 & 62.75 $\pm  0.19$  & $-0.21 \pm$ 0.14 &
253.8 $\pm$ 2.9 & 1.6 \\
\hline
\smlucidEventOr & 1.99 $\pm$ 0.16 & 62.37 $\pm$ 0.16 & 0.13 $\pm$ 0.13 &
825.3 $\pm$ 3.1 & 0.8 \\
%%\hline 
%%\smlucidHitAnd\ & $1.77 \pm 0.23$  & $ 61.75 \pm 0.23 $  & $-0.93 \pm 0.22$  & $1202.7 \pm 11.4$  & 1.8 \\
%%\hline
%%\smlucidHitOr\ & $1.89 \pm 0.17$  & $ 62.39 \pm 0.18 $  & $0.25 \pm 0.48$  & $2336.7 \pm 8.7$  & 1.5 \\
\hline
\smmbtsoneone\ & 2.17 $\pm$ 0.15 & 62.18 $\pm$ 0.16 & 0.30 $\pm$ 0.15 &
1068.9 $\pm$ 3.9 & 0.9 \\
\hline
\smmbtsone\ & 2.11 $\pm$ 0.15 & 62.13 $\pm$ 0.15 & 1.70 $\pm$ 0.20 & 
1207.6 $\pm$ 4.2 & 1.0 \\
\hline
\smmbtsOFF\  & $2.22\pm 0.24 $ & $62.61 \pm 0.27$  & N/A & $1038.0 \pm 3.8$ & 1.5\\
\hline
\smmbtspvtx\  & $2.09\pm 0.21$ & $62.48 \pm 0.25$ & N/A & $1081.0\pm 3.6 $  & 0.9 
\\
\hline
\smchpart\      & $2.27\pm$ 0.36 & $ 62.3 \pm 0.3 $  & N/A & $841.2\pm 4.1$ \
 & 1.1\\
\hline
\smLARtiming\ & $2.50 \pm 0.29$  & $ 62.7 \pm 0.4 $  & N/A & $950.6 \pm 5.4$\
  & 3.0 \\
%\hline
%\smbcmEventAnd\ & $-1.09 \pm 3.30$  & $ 62.37 $ (fixed) & $-0.077 \pm 0.081$ & $2.86 \pm 0.24$  & 0.8 \\
\hline
\smbcmEventOr\  & $1.85 \pm 0.63$  & $ 62.37 $ (fixed) & $0.429 \pm 0.079$ & $85.53 
\pm 0.89$  & 1.2 \\
%\hline
%\smbcmEventXOr\ & $2.35 \pm 0.92$  & $ 62.37 $ (fixed) & $0.224 \pm 0.054$ & $42.21 \pm 0.63$  & 1.2 \\
%\hline
%\smzdcEventAnd\  & $2.93 \pm 0.50 $  & $ 60.86 \pm 0.55 $  & $0.02 \pm 0.05$ & $88.72 \pm 1.1$  & 0.9 \\
\hline
\smzdcA\ & $2.54 \pm  0.25$ & $62.00 \pm 0.27 $ & $0.45 \pm 0.12$ & $368.9 \pm 2.3$  
& 1.1\\
\hline
\smzdcC\      & $2.15 \pm 0.25 $ & $62.38 \pm 0.28 $ & $0.34 \pm 0.12 $ & $355.9 
\pm 2.3$ & 0.8 \\
\hline
\end{tabular}
\caption{Summary of the relevant fit parameters for the Beam Scan~II.
For offline algorithms, the rates have been corrected for trigger prescales.
Because the rates in the BCM were low, the value of $\Sigma$ used for
the BCM was fixed to that obtained from the \lucidEventOr .
}
\label{tab:VdMFits-II}
\end{table*}
}
{\small
\begin{table*}
\centering
\begin{tabular}{|l||c|c|c|c|c|}
\hline
Algorithm & Mean Position & $\Sigma $ &
Background & $R^{MAX}$ & $\chi^2/DOF $\\
 &  $ (\mu m)$ &  $(\mu m)$ &    (Hz) &    (Hz) &  \\
\hline
\multicolumn{6}{|c|}{Horizontal Scan~III} \\
\hline
\smlucidEventAnd & 5.48 $\pm$ 0.26 & 58.94 $\pm 0.19$  & 0.04 $\pm$ 0.13 &
266.8 $\pm$ 3.0 & 1.2 \\
\hline
\smlucidEventOr &  5.66 $\pm$ 0.15 & 58.57 $\pm$ 0.18 & 0.42 $\pm$ 0.10 &
856.8 $\pm$ 3.3 & 2.1 \\
%%\hline
%%\smlucidHitAnd\ & $5.33 \pm 0.26$ & $58.26 \pm 0.36 $  & $0.19 \pm 0.23$  & $1237.8 \pm 10.0$  & 1.4 \\
%%\hline
%%\smlucidHitOr\ & $5.54 \pm 0.15$  & $58.83 \pm 0.19 $  & $1.21 \pm$ 0.31 & $2409.7 \pm 10.8$  & 3.1 \\
\hline
\smmbtsoneone\ & 5.59 $\pm$ 0.14 & 58.88 $\pm$ 0.10 & 0.15 $\pm$ 0.14 &
1102.5 $\pm$ 3.2 & 2.3 \\
\hline
\smmbtsone\ & 5.59 $\pm$ 0.14 & 58.87 $\pm 0.10$  & 1.20 $\pm$ 0.30 &
1244.4 $\pm$ 3.9 & 2.5 \\
\hline
\smmbtsOFF\  & $6.02\pm 0.22$ & $59.05 \pm 0.23$ & N/A & $1074.0\pm 4.0 $   & 0.95 
\\
\hline
\smmbtspvtx\  & $5.95\pm 0.20$ & $59.14\pm 0.23$  & N/A & $1120.0 \pm 3.8$  & 1.4 
\\
\hline
\smchpart\    & $6.03 \pm 0.33$  & $ 59.3 \pm 0.2 $  & N/A & $869.6\pm 4.2\
$  & 1.1 \\
\hline
\smLARtiming   & $6.15 \pm 0.28$ & $ 59.1 \pm 0.3 $  & N/A & $981.7 \pm 6.6\
$  & 1.4 \\
%\hline
%\smbcmEventAnd\  & $4.69 \pm 3.48$ & $58.57 $(fixed)  & $0.05 \pm 0.05$ & $2.40 \pm 0.22$  & 0.63 \\
\hline
\smbcmEventOr\ & $6.36 \pm 0.60$ & $ 58.57 $ (fixed) & $0.23 \pm 0.11$ & $89 \pm 
1$  & 1.25 \\
%\hline
%\smbcmEventXOr\ & $5.74 \pm 0.88$ & $ 58.57 $ (fixed) & $0.10\pm 0.05$ & $43.44 \pm 0.66$  & 1.45 \\
%\hline
%\smzdcEventAnd\  & $5.22 \pm  0.45$  & $ 57.87 \pm 0.27 $  & $-0.05 \pm 0.06$ & $90.2 \pm 0.9$  & 0.8 \\
\hline
\smzdcA\ & $5.38 \pm 0.22$  & $ 59.15 \pm 0.36 $  & $0.28 \pm 0.18$ & $373.6 \pm 
3.1$  & 1.3 \\
\hline
\smzdcC\ & $5.67\pm 0.23$ & $59.01 \pm 0.15 $ & $0.13 \pm 0.10 $ & $366.7 \pm 1.8$ 
& 1.7 \\
\hline
\multicolumn{6}{|c|}{Vertical Scan~III} \\
\hline
\smlucidEventAnd  &  $-0.01 \pm$ 0.27 & 62.21 $\pm$ 0.30 & $-0.03 \pm$ 0.08 &
259.9 $\pm$ 2.9 & 0.9 \\
\hline
\smlucidEventOr & 0.08 $\pm$ 0.16 & 62.06 $\pm$ 0.16 & 0.23 $\pm$ 0.12 &
830.2 $\pm$ 3.1 & 0.8 \\
%%\hline
%%\smlucidHitAnd\ & $-0.15 \pm 0.29$  & $ 61.54 \pm 0.31 $  & $-0.01 \pm 0.41$  & $1209.0 \pm 8.2$  & 1.0 \\
%%\hline
%%\smlucidHitOr\ & $-0.07 \pm 0.17$  & $ 62.23 \pm 0.18$  & $0.64 \pm 0.42$  & $2346.4 \pm 9.2$  & 1.2 \\
\hline
\smmbtsoneone\ & 0.04 $\pm$ 0.15 & 62.09 $\pm$ 0.16 & 0.15 $\pm$  0.15 &
1075.6 $\pm$ 3.9 & 1.2 \\
\hline
\smmbtsone\ & 0.06 $\pm$ 0.15 & 62.09 $\pm$ 0.15 & 1.65 $\pm$ 0.22 &
1214.5 $\pm$  4.2 & 1.1 \\
\hline
\smmbtsOFF\  & $-0.16\pm 0.24$ & $61.45 \pm 0.30$  & N/A& $1056.0 \pm 4.0$  & 1.4 
\\
\hline
\smmbtspvtx\  & $-0.06\pm 0.21$ & $61.83 \pm 0.27$  & N/A & $1102.0 \pm 3.7$  & 
1.4\\
\hline
\smchpart\      & $-0.32\pm$ 0.36 & $ 61.5 \pm 0.3 $  & N/A & $840.6\pm 4.1$\
  & 0.9 \\
\hline
\smLARtiming   & $-0.53 \pm 0.30$  & $ 61.7 \pm 0.5 $  & N/A & $951.1 \pm 6\
.2$  & 3.6 \\
%\hline
%\smbcmEventAnd\  & $-1.75 \pm 4.0$  & $ 62.06 $ (fixed) & $-0.05 \pm 0.06 $ & $2.35 \pm 0.25$  & 0.87 \\
\hline
\smbcmEventOr\ & $0.3 \pm 0.64$  & $ 62.06 $ (fixed) & $0.17 \pm 0.08$ & $86.2 \pm 
1$  & 1.56 \\
%\hline
%\smbcmEventXOr\ & $0.58 \pm 0.93$ & $ 62.06 $ (fixed)  & $0.08 \pm 0.05$ & $41.86 \pm 0.63$  & 0.82 \\
%\hline
%\smzdcEventAnd\  & $0.74 \pm 0.50$  & $ 60.11 \pm 0.45 $  & $0.12 \pm 0.06$ & $88.5 \pm 1.3$  & 1.0 \\
\hline
\smzdcA\ & $-0.04 \pm 0.25$  & $ 62.36 \pm 0.28 $  & $0.17\pm 0.10$ & $367.9 \pm 
2.3$  & 1.2 \\
\hline
\smzdcC\ & $-0.03 \pm 0.25 $ & $62.26 \pm 0.30 $ & $0.42 \pm 0.10 $ & $358.3 \pm 
2.3$ & 0.8 \\
\hline
\end{tabular}
\caption{Summary of the relevant fit parameters for the Beam Scan~III.
For offline algorithms, the rates have been corrected for trigger prescales.
Because the rates in the BCM were low, the value of $\Sigma$ used for
the BCM was fixed to that obtained from the \lucidEventOr .}
\label{tab:VdMFits-III}
\end{table*}
}
\begin{table*}
\begin{center}
\begin{tabular}{|l|c|c|c|}
\hline
Method & Scan Number & $\sigma_{vis}$ 
& $\mathcal{L}_{spec}$ \\
   &   & mb &
($10^{29}\;{\rm cm}^{-2}{\rm s}^{-1}$)\\
\hline
\lucidEventAnd  & 1 & $12.15\pm 0.14 $ & $6.80 \pm 0.08$\\
          & 2 & $12.55\pm 0.10 $  & $4.85 \pm 0.03 $\\
          & 3 & $12.73\pm 0.10 $ & $4.88 \pm 0.03$\\
\hline
\lucidEventOr   & 1 & $39.63\pm 0.32 $ & $6.85 \pm 0.06$\\
          & 2 & $40.70\pm 0.13 $ & $4.88 \pm 0.01$\\
          & 3 & $40.77\pm 0.14 $ & $4.92\pm 0.02$\\
\hline
\mbtsoneone\  & 1 & $51.14\pm 0.39 $ & $6.78 \pm 0.05$\\
          & 2 & $52.59\pm 0.16 $ & $4.87 \pm 0.01$\\
          & 3 & $52.64\pm 0.16 $ & $4.90 \pm 0.02$\\
\hline
\mbtsone\   & 1 & $57.83 \pm 0.43 $ & $6.79 \pm 0.05$\\
          & 2 & $59.47\pm 0.18 $ & $4.89 \pm 0.01$\\
          & 3 & $59.43\pm 0.25 $ & $4.90 \pm 0.02$ \\
\hline
\mbtsOFF\  & 1 & $ 49.28 \pm 0.31$ & $6.76 \pm 0.05$\\
          & 2 & $ 51.64 \pm 0.23$ & $4.87 \pm 0.03$\\
          & 3 & $ 51.29 \pm 0.24  $ & $4.93 \pm 0.03$ \\
\hline
\mbtspvtx\  & 1 & $53.48 \pm 0.29 $ & $6.73\pm 0.05$ \\
          & 2 & $53.64 \pm 0.22 $ & $4.89\pm 0.03$\\
          & 3 & $53.78 \pm 0.23 $ & $4.89 \pm 0.02$\\
\hline
\chpart\      & 1 & $42.61 \pm 0.26 $ & $6.74 \pm 0.05 $  \\
              & 2 & $42.84 \pm 0.21$ & $ 4.91 \pm 0.02 $  \\
              & 3 & $42.93 \pm 0.21 $ & $4.91 \pm 0.03  $  \\
\hline
\LARtiming\   & 1 & $46.43 \pm 0.31$  & $6.78 \pm 0.05  $  \\
              & 2 & $46.98 \pm 0.27$ & $4.91 \pm 0.02 $  \\
              & 3 & $46.63 \pm 0.30$ & $4.91 \pm 0.03 $  \\
%%\hline
%%\zdcEventAnd\ & 2 & $4.36 \pm 0.10 $  & $ 5.10 \pm 0.15 $  \\
%%              & 3 & $4.29 \pm 0.09 $ & $ 5.15 \pm 0.05 $  \\
\hline
\zdcA\        & 2 & $18.12 \pm 0.09$  & $ 4.89 \pm 0.02 $   \\
              & 3 & $17.78 \pm 0.13$  & $ 4.85 \pm 0.04 $  \\
\hline
\zdcC\        & 2 & $ 17.56\pm 0.09$  & $ 4.88 \pm 0.02 $   \\
              & 3 & $ 17.39\pm 0.11$  & $ 4.86 \pm 0.03 $  \\
\hline
\end{tabular}
\end{center}
\caption{Measurements of the visible cross section 
and peak specific luminosity for all algorithms that have been
calibrated using the \vdM\ scan data for each of the three beam scans.  
The uncertainties reported here are statistical only.  
The emittance during Scan~I was
different from that during Scans~II and~III, so the specific luminosity in that
first scan is not expected to be the same.
No results for Scan~I are presented for the ZDC, since the constant fraction 
discriminators used for the ZDC measurements were installed later
in the run. Because the rates in the BCM were low, the value of 
$\Sigma$ used for
the BCM was fixed to that obtained from the \lucidEventOr, so no
measurement of the specific luminosity using BCM data is performed.
}
\label{tab:sigmavisperscan}
\end{table*}

\clearpage
%
%%%%%%%%%%%%%%%%%%%%%%%%%%%%%%%%%%%%%%%%%%%%%%%%%%%%%%%%%%%%%%%%%%%%%%%%%%%%%%%
% Bibliography
%%%%%%%%%%%%%%%%%%%%%%%%%%%%%%%%%%%%%%%%%%%%%%%%%%%%%%%%%%%%%%%%%%%%%%%%%%%%%%
%
% Style file to use with mcite.
% Use atlasstyle with just cite.
\newpage
\bibliographystyle{atlasstylem}
\bibliography{lumiPaper}
\newpage
% ATLAS Collaboration author list for 08-NOV-2010
% Data extracted on 30-Nov-2010 for paperid 44
\begin{flushleft}
{\Large The ATLAS Collaboration}

\bigskip

G.~Aad$^{\rm 48}$,
B.~Abbott$^{\rm 111}$,
J.~Abdallah$^{\rm 11}$,
A.A.~Abdelalim$^{\rm 49}$,
A.~Abdesselam$^{\rm 118}$,
O.~Abdinov$^{\rm 10}$,
B.~Abi$^{\rm 112}$,
M.~Abolins$^{\rm 88}$,
H.~Abramowicz$^{\rm 153}$,
H.~Abreu$^{\rm 115}$
% ...
\bigskip

% ...
$^{10}$ Institute of Physics, Azerbaijan Academy of Sciences, H. Javid Avenue 33, AZ 143 Baku, Azerbaijan\\
$^{11}$ Institut de F\'isica d'Altes Energies, IFAE, Edifici Cn, Universitat Aut\`onoma  de Barcelona,  ES - 08193 Bellaterra (Barcelona), Spain\\
% ...
$^{48}$ Albert-Ludwigs-Universit\"{a}t, Fakult\"{a}t f\"{u}r Mathematik und Physik, Hermann-Herder Str. 3, D - 79104 Freiburg i.Br., Germany\\
$^{49}$ Universit\'e de Gen\`eve, Section de Physique, 24 rue Ernest Ansermet, CH - 1211 Geneve 4, Switzerland\\
% ...
$^{88}$ Michigan State University, Department of Physics and Astronomy, High Energy Physics Group, East Lansing, MI 48824-2320, United States of America\\
% ...
$^{111}$ University of Oklahoma, Homer L. Dodge Department of Physics and Astronomy, 440 West Brooks, Room 100, Norman, OK 73019-0225, United States of America\\
$^{112}$ Oklahoma State University, Department of Physics, 145 Physical Sciences Building, Stillwater, OK 74078-3072, United States of America\\
% ...
$^{115}$ LAL, Univ. Paris-Sud, IN2P3/CNRS, Orsay, France\\
% ...
$^{118}$ Oxford University, Department of Physics, Denys Wilkinson Building, Keble Road, Oxford OX1 3RH, United Kingdom\\
% ...
$^{153}$ Tel Aviv University, Raymond and Beverly Sackler School of Physics and Astronomy, Ramat Aviv, IL - Tel Aviv 69978, Israel\\
% ...
\end{flushleft}

\end{document}